\documentclass[aps,prb,twocolumn,natbib,showpacs,floatfix,superscriptaddress, longbibliography]{revtex4-2}

\usepackage{graphicx}
\usepackage{amsmath,amsthm,amssymb,amsfonts,bbold,bm,color,mathtools}
\usepackage{float}
\usepackage{epstopdf}
\usepackage{hyperref}
\usepackage{enumerate}
\usepackage{wasysym}
\usepackage{url}
\usepackage{dsfont}
\usepackage{braket}
\usepackage{comment} 
\usepackage[capitalize]{cleveref}
\usepackage[normalem]{ulem}
\usepackage{xcolor}
\usepackage{subfigure}
\usepackage{overpic}
\usepackage{svg}
\graphicspath{{figures/}}
\hypersetup{colorlinks=true,linkcolor=blue,citecolor=blue,urlcolor=blue}

\begin{document}
\newcommand{\pv}[1]{\textcolor{black}{#1}}
\newcommand{\jt}[1]{\textcolor{black}{#1}}
\newcommand{\brac}[1]{\langle #1 \rangle}
\newcommand{\para}[1]{\left( #1 \right)}

\title{Probing Pair Density Waves with Twisted Josephson Junctions}

\author{Jefferson Tang}
\affiliation{Department of Physics, University of Connecticut, Storrs, Connecticut 06269, USA}
\author{Pavel A. Volkov}
\affiliation{Department of Physics, University of Connecticut, Storrs, Connecticut 06269, USA}
\begin{abstract}
    We show that twisted interfaces between superconductors can serve as a phase-sensitive platform for the detection and characterization of pair density waves (PDW). In the presence of an in-plane magnetic field, the critical Josephson current of a twisted PDW interface is maximal at a finite field value, determined by the twist angle and the PDW period - an explicit signature of the PDW. The results are robust to variations in junction geometry and can be adapted to certain cases with strong disorder or fluctuations. Their temperature dependence allows to distinguish pure PDW from byproducts of coexistence of superconductivity and charge- or spin- density waves. 
    %Our results establish a phase-sensitive probe of PDW at macroscopic scales via transport measurements.
    % We demonstrate that twisted interfaces between superconductors can serve as a phase-sensitive platform for the detection and characterization of pair density waves (PDW).
    % %pair density wave (PDW) superconductors can serve as study the Josephson effect for twisted junctions between pair density wave (PDW) superconductors. 
    % In the presence of an in-plane magnetic field and nonzero twist, the critical current of such junctions is maximal at a finite field value,  
    % determined by the twist angle and the PDW period, yielding - an explicit signature of the PDW. The results are robust to variations in junction geometry and can be adapted to certain cases with strong disorder or fluctuations. The temperature dependence allows to differentiate between the pure PDW superconductors and states where a PDW arises as a byproduct of coexistence of superconductivity and charge density waves. Our results establish a phase-sensitive probe of PDW correlations at a macroscopic scale via transport measurements.
\end{abstract}
\maketitle

\begin{figure}[t]
    \centering
    \includegraphics[width=1\linewidth]{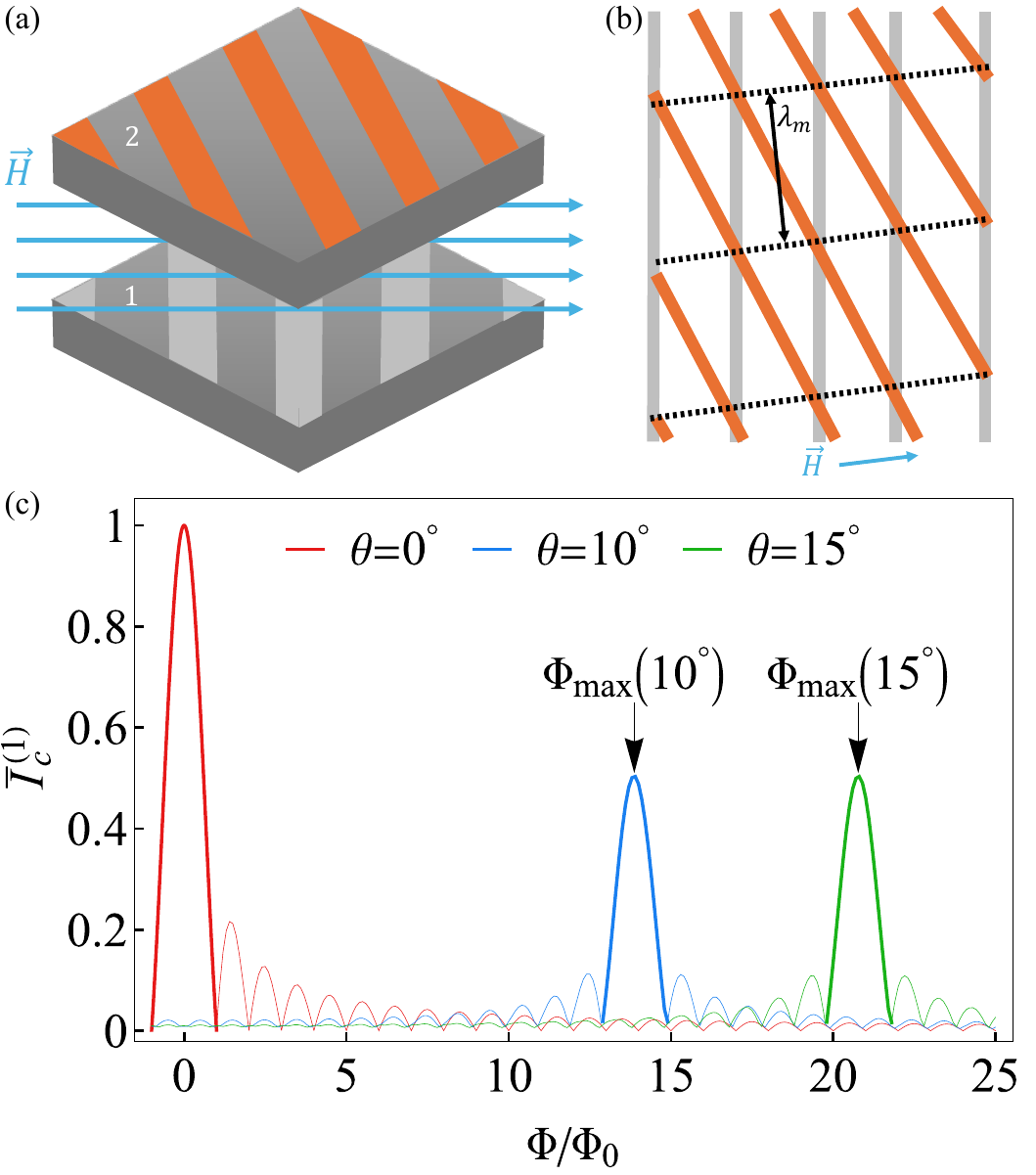}
    \caption{(a) Schematic of the proposed setup. We consider a Josephson junction formed at the interface of two identical bulk twisted pair density wave superconductors (gray/orange stripes) in the presence of an in-plane magnetic field $\vec{H}$.
    At the interface (b), the overlapping twisted PDWs form a moir\'e pattern with period $\lambda_m$ creating a quasiperiodic phase mismatch between the order parameters. (c) Normalized critical current $\overline{I}^{(1)}_c$, Eq. \eqref{rect_current} of as a function of total magnetic flux $\Phi$ through the junction for field oriented as in (b). At a nonzero twist angle $\theta$, $\overline{I}^{(1)}_c$ shows a maximum at a finite $\Phi=\Phi_{max} \propto \lambda_m^{-1}$ \eqref{eq:phimax}, that provides a direct evidence of PDW. The junction width \pv{$L/\lambda_{\text{PDW}}\approx80$} was used in this plot.    
    }
    \label{fig:cartoon}
\end{figure}

{\it Introduction---} Pair-density wave (PDW) is a superconducting state in which Cooper pairs have a nonzero momentum, resulting in a spatially-modulated order parameter that breaks translational symmetry \cite{Agterberg2020}. While originally
such form of pairing has been proposed to be induced by magnetic field \cite{Larkin1964,Fulde1964}, more recent studies suggest that PDW can also occur due to strong electronic correlations \cite{Agterberg2020,deveraux2024,Huang2022,hongyao_2025}. A number of mechanisms for the PDW formation have been proposed, including competition or coexistence with other (intertwined) orders \cite{himeda2002,Berg2007,Berg2009newphy,kopp_2011,sotogarrido2014,vafek2014,chubukov2015,intertwined_2015}, Amperean pairing 
\cite{Lee2014,kargarian_2016} and others \cite{yiming_2023,Setty2023,kunyang2023,Berg2010,cpt_2022,yahui2022,zhaoyu2025,bitan2010,Venderley2019,zhaoyu2022,Shaffer2023,jiang2024,sarma2025,hongyao_2025,fcwu2023,Scammell2023,Schwemmer2024,Wu2023,Yao2025,gil2025,cho_2012,Zhou2022,zhaoyu2024,slagle_2020,wu2003_prl,sheng_2023,wang2025negativeroutepairdensity,scalettar2025}.
% A number of mechanisms for PDW formation have been proposed, including competition or coexistence with other (intertwined) orders \cite{himeda2002,Berg2007,Berg2009newphy,kopp_2011,sotogarrido2014,vafek2014,chubukov2015,intertwined_2015}, Amperean pairing 
% \cite{Lee2014,kargarian_2016}, momentum structure of the interactions
% \cite{yiming_2023,Setty2023,kunyang2023}, Kondo physics \cite{Berg2010,cpt_2022,yahui2022,zhaoyu2025}, valley physics in hexagonal systems \cite{bitan2010,Venderley2019,zhaoyu2022,Shaffer2023,jiang2024,sarma2025,hongyao_2025}, van Hove singularities \cite{fcwu2023,Scammell2023,Schwemmer2024,Wu2023,Yao2025,gil2025} , 
% interplay with band topology and quantum geometry \cite{cho_2012,Zhou2022,zhaoyu2024} as well as 
% moir\'e effects in twisted bilayers \cite{slagle_2020,wu2003_prl,sheng_2023}. 
Experimental signatures of PDWs at zero field have been observed to date in the cuprate \cite{Hamidian2016,Du2020}, heavy fermion \cite{hf1,hf2}, and iron-based superconductors \cite{fujita2023,Liu2023} and, more recently,  UTe$_2$ \cite{Gu2023,Aishwarya2023,Aishwarya2024}, kagome materials \cite{Chen2021,Deng2024}, and transition metal dichalcogenides \cite{Liu2021,Devarakonda2024,Cao2024}. 

The overwhelming majority of evidence for PDW arises from scanning tunneling microscopy (STM) \cite{Hamidian2016,Du2020,fujita2023,Liu2023,Gu2023,Aishwarya2023,Aishwarya2024,Chen2021,Deng2024,Liu2021,Cao2024}, which probes PDW locally and within a limited field of view. 
%The latter may be an important for the scenarios where PDW is fluctuating \cite{Lee2014,Setty2023} may lack long-range order. 
Moreover, phase-sensitive measurements require  superconducting tip (Josephson STM)  \cite{Hamidian2016,Du2020,Liu2021,Chen2021,Deng2024}; in the future, two-electron photoemission may have a similar capability \cite{2earpes}. Other works rely on indirect probes that are not phase-sensitive, such as transport \cite{hf1,Devarakonda2024}, coupling to charge- or spin- modulations \cite{hf2}, or collective mode spectroscopy \cite{Wu2025}. Development of phase-sensitive probes of PDW at macroscopic scales is thus required for the understanding the PDW as a macroscopic coherent state of matter. A potential unexplored direction is suggested by the recent experimental and theoretical advances in twisted interfaces between unconventional superconductors (Ref. \onlinecite{pixley2025} and references therein). Transport properties of such interfaces containing phase-sensitive information about the order parameter \cite{klemm2005phase,xiaoberg_2023,yuan2024phase}; however, the properties of twisted PDW interfaces have not been studied to date.
%. Recent theoretical proposal suggest the applications of this technique to probe unconventional van der Waals superconductors  \cite{xiaoberg_2023} or using edge effects to extract the full orientation dependence of the order parameter \cite{yuan2024phase}. However, the properties of twisted PDW interfaces have not been studied to date.

%An unexplored route of probing the PDW state has been opened up by recent advances in the construction of van der Waal materials, as it is possible to construct scanning probes sensitive to the behavior of electrons at a macroscopic scale \cite{Inbar2023}. 

%Josephson effect - sensitive to pairing.

%In the presence of twist, the PDWs form a moir\'e pattern at the interface (Fig. \ref{fig:cartoon} (b)), frustrating the Josephson coupling between the order parameters across the int.

Here we demonstrate that twisted interfaces between bulk pair density wave superconductors (Fig. \ref{fig:cartoon} (a)) provide a platform for phase-sensitive probing of the PDW state that can be implemented using current experimental capabilities. In the presence of a magnetic field in the junction plane, the critical current $I_c$ (Fig, \ref{fig:cartoon} (c)) develops oscillations with a maximum at a finite, twist-dependent field value, in contrast to conventional superconductors where the peak is expected at zero field \cite{barone1982}. Qualitatively, the peak arises when an integer number of flux quanta threads through a unit cell of moir\'e superlattice formed by the twisted PDWs at the interface (Fig. \ref{fig:cartoon} (b)). Its presence is robust to variations in junction geometry and its temperature dependence further allows one to distinguish between pure PDWs and PDWs arising from coexistence of other order parameters. We also suggest that the Josephson current statistics at a fixed phase difference can probe signatures of certain types of disordered or fluctuating PDWs. Our results establish an experimentally accessible route for probing and characterizing PDW states at macroscopic scales.

{\it Model---} To illustrate the main principles of our proposal, we will focus on the case of a twisted interface between identical stripe (single-$\vec{q}$) PDWs, depicted in Fig. \ref{fig:cartoon} (a). %We consider the order parameter physics of the single-$\vec{q}$ PDW superconductor and discuss the generalization later. 
%We begin with two twisted flakes of identical PDW superconductors. 
We further disregard the effects of unconventional internal symmetry of the order parameter (such as d-wave), as it does not affect the coupling across the interface at sufficiently low twist angles \cite{Tummuru-Franz-PRB-2022,volkov_2025}.
%For low twist angles we can neglect the unconventional symmetry of the order parameter, such as d-wave. 
We write the order parameter of the individual layers as \cite{Larkin1964,Agterberg2008,Agterberg2020,Liu2021}
\begin{equation}
    \Delta_j(\vec{r}) = |\Delta_{q}|e^{i\phi_j}
    \cos(\vec{q}_j\cdot\vec{r}+\eta_j)
    %\left(e^{i(\vec{q}_j\cdot\vec{r}+\eta_j)} + e^{-i(\vec{q}_j\cdot\vec{r}+\eta_j)}\right),
    \label{eq:pdwOP}
\end{equation}
where $j=1,2$ is the layer index [See Fig.\ref{fig:cartoon}] and $|\Delta_q|$ is the magnitude of the order parameter, which is identical in both layers. The phases $\phi_{1,2}$ and $\eta_{1,2}$ correspond to the spontaneous breaking of the $U(1)$ and translational symmetry, respectively.
%The breaking of the global $U(1)$ and translational symmetry occurs due to spontaneously chosen phases $\phi$ and $\eta$. 
The center-of-mass momentum of the Cooper pairs is given by $\hbar\vec{q}_{1,2}$, leading to a spatial modulation of the order parameter with period $\lambda_{PDW} = \frac{2\pi}{q}$
%hich also describes the spatial modulation of the order parameter 
\cite{Larkin1964,Fulde1964,Agterberg2008,Agterberg2020,Berg2009newphy,Berg2009PRB}. In the presence of interflake twist angle $\theta$, $\vec{q}_{1,2}$ have the same magnitude but are rotated with repect to one another by $\theta$.

We now consider the Josephson current across the twisted interface. To derive the expression, we follow a phenomenological approach \cite{pixley2025,can2021high,volkov_diode,tang2025}, starting with an expansion of the interfacial free energy $F_{12}[\Delta_1(\vec{r}),\Delta_2(\vec{r})]$ in the order parameters. This is valid either close to $T_c$, or for weak tunneling across the interface \cite{Tummuru-Franz-PRB-2022,volkov_2025}. The Josephson current is then found as $I =\partial F_{12}/\partial \phi_{12}$ \cite{KupriyanovRev}, where $\phi_{12}=\phi_{1}-\phi_2$ is the superconducting phase difference. Assuming that the coupling between the order parameters is local on the scale of the PDW period $\lambda_{PDW}$, we get in the lowest order in $\Delta_1(\vec{r}),\Delta_2(\vec{r})$:
%The low-energy behavior of the junction can be captured by a local Josephson coupling term, $F_{12}$, at zero twist and the superconducting phase difference $\phi_{12}=\phi_{1}-\phi_2$. Phase sensitive measurements can be then carried out by measuring the Josephson current, $I =\partial F_{12}/\partial \phi_{12}$ \cite{KupriyanovRev}. We will first focus on the first harmonic current, as it is the lowest order term in the weak coupling limit or for temperatures close to $T_c$:
\begin{align}
    I^{(1)} &=\ A \int d\vec{r}  \ \text{Im}\Big[\Delta_1^*(\vec{r})\Delta_2(\vec{r})-\Delta_1(\vec{r})\Delta_2^*(\vec{r})\Big], 
    \label{eq:I1}
\end{align} 
where $A$ is a constant and the integral is carried over the junction area. Substituting $\Delta_{1,2}(\vec{r})$ from Eq. \eqref{eq:pdwOP} into Eq. \eqref{eq:I1} one obtains:
\begin{equation}\label{first_harmonic} 
    I^{(1)} = \int d\vec{r} \ 
    J_{c1}
    [\cos(q_-x+\eta_{12})
    +
    \cos(q_+y+\eta_1+\eta_2)]\sin\Big(\phi_{12}\Big),
\end{equation}
where $\vec {q}_\pm = \vec{q}_1\pm\vec{q}_2$ with $x,y$ chosen to be along $q_-,q_+$, respectively, and $\eta_{12} = \eta_1-\eta_2$. At zero twist, $q_-=0, q_+=2q_{PDW}$, the former resulting in a nonzero $I^{(1)}$, proportional to the interface area.  At finite twist, both terms in Eq. \eqref{first_harmonic}  oscillate in space as $|\vec{q}_{\pm}| =q_{PDW}\sqrt{2(1\pm\cos(\theta))}$. Nonzero $q_-$ in particular describes the moir\'e modulation of the relative phase between twisted PDWs (Fig. \ref{fig:cartoon}(b)) with period $\lambda_m = \frac{2 \pi}{q_-}$, that diverges as $\theta \to 0$ \footnote{For twist angles larger than $90^\circ$, the role of $q_+$ and $q_-$ as defined above, interchanges.}. We highlight that this moir\'e modulation is a characteristic feature of twisted PDWs. While conventional superconductivity is also modulated by the atomic lattice of the material, the order parameter is still expected to contain a uniform component that will not transform under twist, in contrast to PDW.
%\pv{TBD: argument against s-wave moire}
As a result, the Josephson current density (integrand of Eq. \eqref{first_harmonic}) contains at low $\theta$ a slowly oscillating component that nonetheless integrates to zero. It is known however \cite{barone1982}, that the critical Josephson current in presence of magnetic field $H$, $I_c(H) =\underset{\phi_{12},\eta_{12}}{\max}[I(H)]$, probes the Fourier transform of the critical current density, suggesting a route to detect the PDW moir\'e pattern.

Let us start by considering the effect of an in-plane field along $y$ (Fig. \ref{fig:cartoon} (b)). We assume that the external field is sufficiently small, such that the bulk of the two PDW flakes remains in the Meissner state. For a short junction, where the Josephson penetration depth $\lambda_J$ \cite{barone1982,owen1967} \footnote{which depends on material parameters, such as the London penetration depth, but grows with decreasing Josephson coupling, i.e. weak junctions are can be usually considered short.}, is much larger than the junction dimension along $x$ which we call $L$, magnetic field modifies $\phi_{12}$, Eq. \eqref{first_harmonic}, as follows \cite{tinkham2004,barone1982}:
%The physics of Josephson junction in the presence of an in-plane magnetic field [See Fig. \ref{fig:cartoon}(a)] is determined by the relative size of the junction and the Josephson length, $\lambda_J$, the latter of which corresponds to the magnetic field penetration depth into the junction \cite{barone1982,owen1967}. We assume that the external field is sufficiently small, such that the bulk of the two PDW flakes remains in the Meissner state. Further, we will focus on the limit where $\lambda_J$ is smaller than the dimensions of the junction [see \cite{supp} for discussion of long junctions]. We first consider an in-plane magnetic along the $x$ direction (we consider the effect of field orientation later), which induces a spatial gradient in the superconducting phase difference of Eq. \eqref{first_harmonic} \cite{tinkham2004,barone1982}:
\begin{equation}\label{phase_dif_mag}
    \phi_{12}(x) =q_Hx+\phi_0,
\end{equation}
where $q_H=(2\pi d/\Phi_0)H$, $d$ is the effective junction thickness \cite{barone1982,volkov_2025}, $\Phi_0 = \frac{\pi \hbar c}{|e|}$ the flux quantum, and $\phi_0$ a constant. For a rectangular junction with dimensions $L$ and $W$ along $x$ and $y$, respectively, the integral over the area of the junction in Eq. \eqref{first_harmonic} can be carried out explicitly. At low twist angles where $q_-\ll q_+$, the second term (with $q_+$) can be neglected for $q_H \ll q_+$.
%Because we are interested the low angle limit where $\lambda_m\gg1/q_+$, we will ignore the $q_+$ term. We stress that while it can be detected (as we show later), we will focus on terms generated by the PDW moire superlattice, $\lambda_m$. 
Under these approximations, the first harmonic current of a rectangular junction, which we normalize by the critical junction field at zero twist and field $I_c(0)= J_{c1} LW$ takes the form:
\begin{align}\label{rect_current}
    \bar{I}^{(1)}\left(\frac{\Phi}{\Phi_0},\eta_{12},\phi_{12}\right) &\approx \frac{\sin\left(\varphi_{+}\right)}{2}\text{sinc}\left(\pi \frac{\Phi_{max}+\Phi}{\Phi_0}\right)  \nonumber \\
    & + \frac{\sin\left(\varphi_{-}\right)}{2}\text{sinc}\left( \pi \frac{\Phi_{max}-\Phi}{\Phi_0}\right)
\end{align}
where $\varphi_{\pm}=\eta_{12}\pm\phi_{12}$, $\Phi = d H L$ is the magnetic flux through the junction, and $\Phi_{max}$ is given by:
\begin{equation}
    \Phi_{max}(\theta)=
    %2 \frac{2 \pi L \sin \frac{\theta}{2}}{\lambda_{PDW}}\Phi_0         =
     \frac{L}{\lambda_{m}(\theta)}\Phi_0,
\label{eq:phimax}
\end{equation}
where $\lambda_m(\theta) = \lambda_{PDW}/(2 \sin \theta/2)$ is the PDW moir\'e period. In Fig. \ref{fig:cartoon}(c) we present the critical current, obtained by maximizing Eq. \eqref{rect_current} with respect to $\eta_{12},\phi_{12}$, normalized by its maximal value at $\theta=0$. While at $\theta=0$, $\overline{I}_c^{(1)}(\Phi)$ shows a standard Fraunhofer pattern-like behavior \cite{barone1982}, increasing the twist angle $\theta$ shifts the peak $\overline{I}_c^{(1)}(\Phi)$ to a finite flux value, well captured by Eq. \eqref{eq:phimax}. The condition $\Phi=\pm\Phi_{max}$ implies that there is a $\Phi_0$ flux penetrating the junction for each PDW moir\'e period. For the peaks in $\overline{I}_c^{(1)}(\Phi)$  to be clearly resolved, one finds from Eq. \eqref{rect_current} that $\Phi_{max} \gtrsim \Phi_0$ is required, which translates to $L\gtrsim \lambda_m(\theta)$. The field strength corresponding to the critical current peak, $H_{max} = 2 \frac{\Phi_0}{d \lambda_m(\theta)}$, decreases with $\theta$ and, in principle, can be as small as $\sim \frac{\Phi_0}{d \lambda_J}$ , for the present result to hold.

The results above have been obtained using a number of simplifying assumptions; we argue now that the appearance of twist-dependent finite-field peaks in the critical current is a generic feature of PDW twist junctions. First of all, they appear independent of the junction geometry. Indeed, the finite field peak arises from the compensation between $q_H$, Eq. \eqref{phase_dif_mag}, and $q_-$ in Eq. \eqref{first_harmonic}. This results in a constant integrand, yielding result proportional to the junction area. The dependence on the geometry thus arises only from the oscillatory terms, and may thus affect the shape of the peak at $\Phi\neq \Phi_{max}$, but not its value at $\Phi=\Phi_{max}$. In the supplementary material, we demonstrate that for a circular junction the results are indeed very similar.

%It is important to note that this result is independent of the junction geometry in the $Lq_-\gg$ limit, as the contribution from mismatch in the modulations in Eq. \eqref{first_harmonic} near the boundaries of the junction is much smaller than the bulk and can be ignored [see \cite{supp} for the example of a circular junction].

\begin{figure}[h]
    \centering
    \includegraphics[width=1\linewidth]{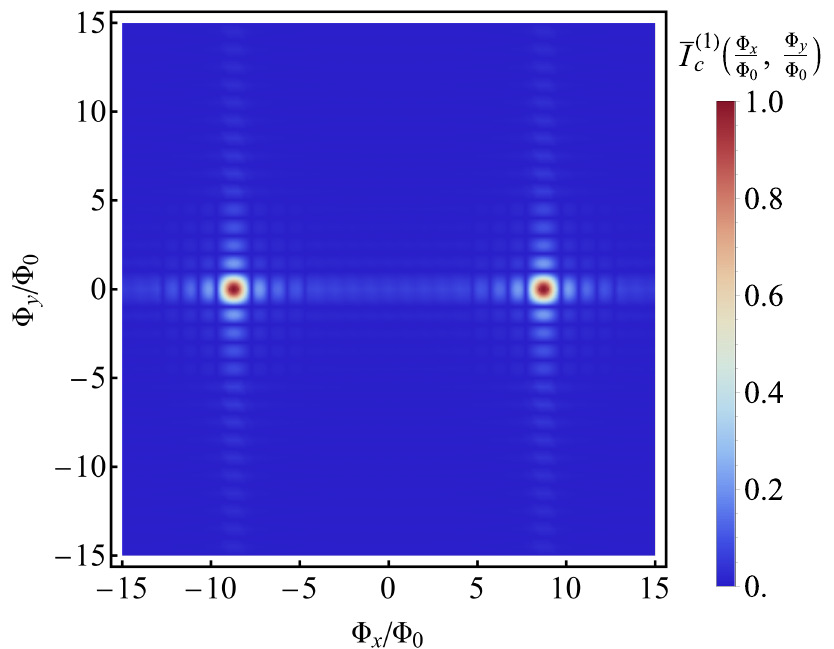}
    \caption{Density plot of the critical Josephson current, Eq. \eqref{field_orientation}, for varied magnetic flux and field orientation. Dependence shown in Fig. \ref{fig:cartoon} (c) corresponds to a cut at $\Phi_y=0$. \pv{We take $L/\lambda_m=8.71$.}
    %$\Phi_{max}(10^\circ)=17.43\Phi_0$. 
    }
    \label{fig:vary_orientation}
\end{figure}

Furthermore, Fig. \ref{fig:cartoon} (c) we assumed a fixed orientation of $\vec{H}$ along the moir\'e modulation (Fig. \ref{fig:cartoon} (b)). In an experimental setting, the direction of the PDW wave-vectors is not known. For an in-plane field oriented in an arbitrary direction, the phase difference is instead $\phi_{12}(x,\alpha) =q_H\Big(\cos(\alpha)x + \sin(\alpha)y\Big)+\phi_0$ \cite{barone1982,tinkham2004}. Substituting $\phi(x,\alpha)$ into Eq. \eqref{first_harmonic} and integrating, we find for a rectangular junction:
\begin{align}\label{field_orientation}
    \bar{I}^{(1)}\para{\frac{\Phi_x}{\Phi_0},\frac{\Phi_y}{\Phi_0}}=\bar{I}^{(1)}\para{\frac{\Phi_x}{\Phi_0}}\frac{\sin\para{\pi\Phi_y/\Phi_0}}{\pi\Phi_y/\Phi_0}
\end{align}
where $\Phi_x=\Phi\cos(\alpha)$ and $\Phi_y=\Phi\sin(\alpha)$. The modulation of $\phi_{12}$ along the y ($\vec{q}_+$) direction in Eq. \eqref{field_orientation} causes the critical current to decay in presence of $\Phi_y\neq 0$. In Fig. \ref{fig:vary_orientation} we present the critical current calculated using Eq. \eqref{field_orientation} as a function of $\Phi_x,\Phi_y$. One observes that while the strongest peaks occur along the $\Phi_y=0$ direction, there are notable peaks at $\Phi_x \approx \Phi_{max}$ for other values of $\Phi_y$ as well. Thus, performing field scans of $I_c(\vec{H})$ with a few field orientations of $\vec{H}$ should be sufficient to distinguish the PDW signatures in an experiment.

%Above we have used several simplifications which may not hold in particular experiments, which we will address now. To begin, 

\begin{figure}[h]
    \centering
    \includegraphics[width=0.95\linewidth]{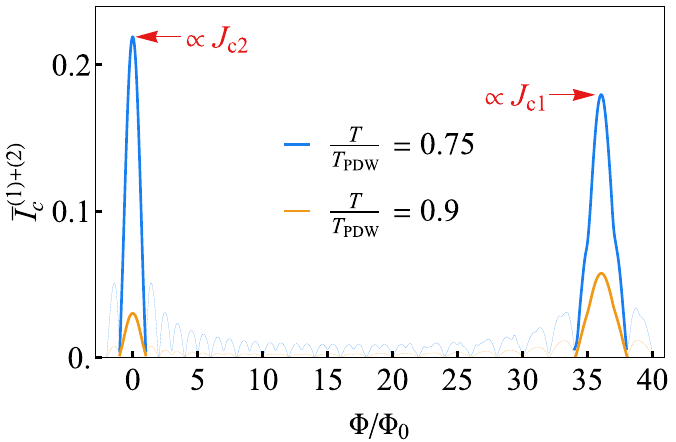}
    \caption{Critical Josephson current of a twist PDW junction in presence of both first and second harmonic (Eq.\eqref{first_harmonic} and Eq. \eqref{second_harm}) for varied magnetic flux at two different temperatures. %Strictly speaking, the Ginzburg-Landau temperature dependence does not hold at temperatures far away from from $T_c$.     However, we plot low temperature values for demonstration purposes. 
    The critical currents are normalized by its value at $\Phi=0$ and $\theta=0^\circ$. To highlight the effects of temperature dependence $T_c$, we take dominant second harmonic $J_{c1}/J_{c2} (T=0)=1/6$ (possible in presence of disorder, see text) and $L/\lambda_m= 36.07$.}
    \label{fig:1st+2nd_harm}
\end{figure}

Finally, the expression of the critical current, Eq. \eqref{eq:I1}, represents only the leading term in the expansion in powers of the order parameter. The next order term allowed by 
%subleading term, or the second harmonic current, that preserves 
the $U(1)$ symmetry is given by \cite{Berg2007,can2021high,pixley2025}:
\begin{align}\label{second_harm}
    I^{(2)} &\equiv\int d\vec{r} \ B \ \text{Im}\left[\Delta^{*2}_1(\vec{r})\Delta^{2}_2(\vec{r})-\Delta^2_1(\vec{r})\Delta^{*2}_2(\vec{r})\right] \nonumber \\
    &\approx\int d\vec{r} \ J_{c2}\sin(\phi_{12})\left[\cos(2q_-x+2\eta_{12})+2\right]
\end{align}
where $B$ is a constant and we have omitted the terms containing $q_+$ for the same reasons as Eq. \eqref{first_harmonic}. The main difference from Eq. \eqref{first_harmonic} is the presence of a constant, non-oscillatory term under the integral in  Eq. \eqref{second_harm}. In presence of a magnetic field, this term results in 
%The constant term in Eq. \eqref{second_harm}  can support 
a zero field peak of $I_c(H)$ proportional to $J_{c2}$ [See Fig. \ref{fig:1st+2nd_harm}], similar to what is expected in a conventional superconductor. Phenomenologically, this peak can also be larger than the one at $\Phi=\pm \Phi_{max}$ for $J_{c2} \gtrsim J_{c1}$ (see also the discussion of a disordered PDW below). However, this situation can be distinguished from the conventional case by its temperature dependence. We will adopt the Ginzburg-Landau temperature dependence, $|\Delta_{q}|(T)=|\Delta_q|(0) \sqrt{1-T/T_{PDW}}$, for the superconducting order parameters. In that case, we obtain $I^{(1)}(T) \propto |\Delta_{q}|^2(T) \propto (1-T/T_{PDW})$ from Eq. \eqref{eq:I1} and $I^{(2)}(T) \propto |\Delta_{q}|^2(T) \propto (1-T/T_{PDW})^2$ from \eqref{second_harm}. Moreover, for conventional superconductors, the zero-field critical current should decrease as $1-T/T_{c}$\cite{amb_bar_1963}, in contrast to the PDW case.

\begin{figure}[h]
    \centering
    \includegraphics[width=0.95\linewidth]{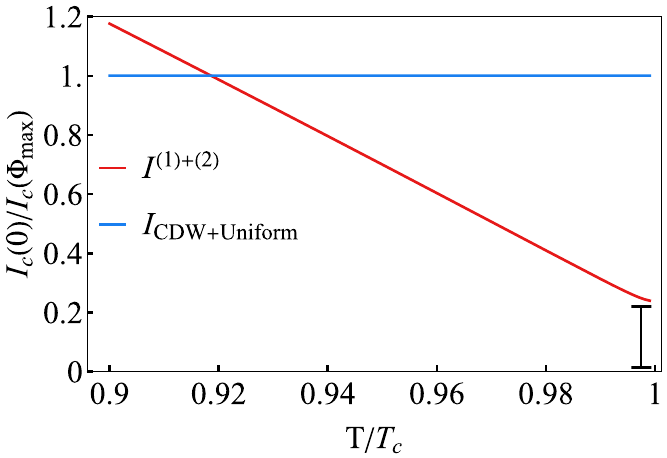}
    \caption{The ratio of the finite field peak to the zero field peak close to $T_c$ for intrinsic PDW (red, see Fig. \ref{fig:1st+2nd_harm}) and ordinary superconductivity coexisting with CDW (blue, Eq. \eqref{cdw_op}). Two scenarios show qualitatively different temperature dependence. We take 
    $L/\lambda_m = 55.27$ %$\Phi_{max}(20^\circ)=27.64\Phi_0$     to demonstrate both the saturation and asymptotic behavior of the PDW peak ratios
    %    limit for a finite sized junction, as a result, the PDW peak ratio saturates near $T_c$, which we explain in the text. We multiplied the ratio for the pure PDW case by a factor of $10$ for better comparison of the two scenarios.
    , $J_{c1}/J_{c2}=1$ at $T=0$, $|\rho_k|^2=\beta=1$, and the field is along y as in Fig. \ref{fig:cartoon} (b).}
    \label{fig:peak_ratio}
\end{figure}

This behavior is illustrated in Fig. \ref{fig:1st+2nd_harm}, where on approaching $T_c$ the finite-field peak becomes dominant, even if it is smaller at low $T$. To elaborate on this, we show the ratio of the zero-field and finite-field peaks as a function of temperature in Fig. \ref{fig:peak_ratio}, red line, calculated for a finite junction size $L/\lambda_m$. The behavior is consistent with the $I^{(2)}(T)/I^{(1)}(T)\propto (1-T/T_{PDW})$, apart from the immediate vicinity of $T_c$. There, the zero-field peak appears due to the finite size effects from the dominant $J_{c1}$. As a result, the peak ratio saturates at the value $2\sin(\pi\Phi_{max}/2\Phi_0)/(\pi\Phi_{max}/2\Phi_0)$.

%In the pure PDW case when $L\rightarrow \infty$, the finite field peak originates from the first harmonic current, while the zero field peak originates from the second harmonic current, as a result of which $\bar{I}_c(0)/\bar{I}_c(\Phi_{max}/\Phi_0)\propto T_c-T$. 

%Thus, the first-harmonic effects should be always dominant near $T_{PDW}$. 

%We can see from Fig. \ref{fig:1st+2nd_harm} that the zero field peak of the PDW case decreases as $(1-T/T_{PDW})^2$, where $T_{PDW}$ is the critical temperature of the PDW system. In contrast, the peak in the conventional case to decrease as $1-T/T_{SC}$, as $I_c\propto |\Delta_0|^2$, where $|\Delta_0|$ is the magnitude of its order parameter. We can also seem from Eq. \eqref{second_harm} that the second harmonic current contains a $2\Phi_{max}$ peak. However, close to the transition temperature, the peak scales as $(1-T/T_{PDW})^2$. 

{\it  Disordered and fluctuating PDWs:} Disorder and inhomogeneity in the materials can affect PDWs, most profoundly by pinning the PDW modulation  \cite{Chen2022,wen2025}. Indeed, disorder can couple directly to the $2 q$ charge density wave, that would be generically induced by PDWs \cite{Agterberg2020}, such that the phase $\eta$ becomes disordered. Alternatively, there exist proposals of intrinsic strong fluctuations in PDWs \cite{Lee2014,Setty2023} and there is a possibility of disordering the $\eta$ phase by proliferating dislocations \cite{Agterberg2020}. We argue that even if $\eta$ is disordered, twisted PDW junctions can still yield evidence of the PDW state. We will assume, however, that $\phi$ is ordered, such that the system is still superconducting. Rather than focusing on the critical current as before, we consider a different experimental setup - specifically, we study the current $I^{(1)}$, Eq. \eqref{first_harmonic}, across the junction for a fixed superconducting phase difference $\phi_{12}$. This can be achieved, for example, by integrating the PDW twist junction in a superconducting loop and threading a magnetic flux through the loop (but not the junction itself), as in an rf- (single-junction) SQUID \cite{tinkham2004}.

We assume that the phase $\eta$ is disordered, such that $\langle e^{i \eta(x) }\rangle = 0$, but that  $\langle e^{i (\eta(\vec{r}) -\eta(\vec{r}') }\rangle $ is nonzero and decays with distance over a correlation length $\xi$. For concreteness, we choose $\langle e^{i (\eta(\vec{r}) -\eta(\vec{r}') }\rangle  = e^{-\frac{|\vec{r}-\vec{r}'|^2}{\xi^2}}$. It follows that $\langle I^{(1)} (\Phi, \eta(x), \phi)\rangle_{\eta(x)} = 0$. Note that this does not imply the vanishing of the Josephson current for each $\eta(x)$ realization, but rather the average. We note also, that the constant part of the second-harmonic critical current density, Eq. \eqref{second_harm}, is immune to this type of disorder, suggesting that $J_{c2}>J_{c1}$ case may indeed be realized in disorder PDW junctions. On the other hand, the average $\langle \left[I^{(1)} (\Phi, \eta(x), \phi)\right]^2\rangle_{\eta(x)}$ does not vanish and for flux close to $\Phi_{max}$ takes the form:
\begin{equation}
\langle \left[I^{(1)} (\Phi, \eta(x), \phi)\right]^2\rangle_{\eta(x)}
      \left. \approx \right|_{q_H \approx \pm q}
     \pi \xi^2 L W
     \frac{J_{c1}^2}{4}
     e^{-\frac{\xi^2 (|q_H|-q)^2}{4}}.
     \label{eq:disord}
\end{equation}
The result shows that there are still finite-field peaks to be observed, with amplitude extensive in the system size.

{\it Coexistence of uniform superconductivity and density waves---} PDWs can also arise as a byproduct of coexistence between uniform superconductivity and charge- or spin- density waves \cite{Agterberg2020,Liu2021}. Our proposal can also distinguish this scenario from the intrinsic PDW. We begin by considering a PDW-like order arising from coexistence of CDW and uniform superconductivity. %, where $T_{CDW} > T_{sc}$. The 
In a Landau expansion, the PDW arises first from a product of the superconducting order parameter and that of the CDW \cite{Agterberg2020}, leading to
%We write the order parameter of the individual layer as
\begin{equation}\label{cdw_op}
    \Psi_{j}(\vec{r})=|\Delta_{0}|e^{i\phi_{j}}\left[1+\beta |\rho_{k}|\cos(\vec{k}_{j}\cdot\vec{r}+\varphi_{j})\right]
\end{equation}
where $j=1,2$ is the layer index, $\beta$ is the relative strength of the two orders, and $|\Delta_0|$ and $|\rho_k|$ are the magnitude of the superconducting and CDW order parameters, respectively, identical in both layers. The phases $\varphi_{1,2}$ corresponds to the breaking of the translational symmetry from the formation of a charge-density order. Similar to Eq. \eqref{eq:pdwOP}, the order parameter modulates with period $\lambda_{CDW} = 2\pi/k$. 

The Josephson current can then be calculated following the same steps as in the pure PDW case, Eqs. (\ref{eq:I1},\ref{first_harmonic}). The crucial difference, however, is that a non-modulated component of the Josephson current density will appear, leading to a peak in the critical current at zero field. Therefore, the system will exhibit two peaks in the experimental setup of Fig. \ref{fig:cartoon} (a): at zero field and a finite field, where $\Phi=\Phi_m$, where the latter arises from the CDW moir\'e pattern.
%Even at the first harmonic, there is a dominant zero field peak from the coupling of the superconducting order parameter in Eq. \eqref{cdw_op}. We can  define a moire superlattice associated with the interference of the CDW. We have shown in Eq. \eqref{rect_current} that due to the PDW moire superlattice, which generates modulation in the supercurrent density [Fig. \ref{fig:cartoon}(b)], it is possible for the maximal Josephson critical current to occur at finite fields. This is also the case when a uniform superconducting order parameter couples to a CDW order. 
Substituting Eq. \eqref{cdw_op} into Eq. \eqref{eq:I1}, we find the the magnitude of the zero and finite field peaks to be $\bar{I}(0)_c\propto |\Delta_0|^2$ and $ \bar{I}_c(\Phi_{max}/\Phi_0)\propto |\Delta_0|^2|\rho_k|^2$. 

%consistent with the breaking of time-reversal and translational symmetry is of the same form as Eq. \eqref{first_harmonic} and for the same reason, we will ignore contributions from terms containing $\vec{k}_+$ (which we define similarly as in the PDW case in Eq. \eqref{first_harmonic}) and $\vec{k}$. Even at the first harmonic, there is a dominant zero field peak from the coupling of the superconducting order parameter in Eq. \eqref{cdw_op}. We can  define a moire superlattice associated with the interference of the CDW. We have shown in Eq. \eqref{rect_current} that due to the PDW moire superlattice, which generates modulation in the supercurrent density [Fig. \ref{fig:cartoon}(b)], it is possible for the maximal Josephson critical current to occur at finite fields. This is also the case when a uniform superconducting order parameter couples to a CDW order. Substituting Eq. \eqref{cdw_op} into Eq. \eqref{eq:I1}, we find the the magnitude of the zero and finite field peaks to be $\bar{I}(0)_c\propto |\Delta_0|^2$ and $ \bar{I}_c(\Phi_{max}/\Phi_0)\propto |\Delta_0|^2|\rho_k|^2$. 

At a first glance,two strong peaks also appear for pure PDWs in the presence of higher-order interface coupling terms, see Eq. \eqref{field_orientation} and Fig. \ref{fig:1st+2nd_harm}. However, the temperature dependence of the peaks in two scenarios is qualitatively different. In Fig. \ref{fig:peak_ratio}, we compare the relative size of the finite and zero field peak as a function of temperature for the two scenarios. For the coexistence case, we assume $T_{CDW}>T_{SC}$ (as in, e.g., TMDs \cite{Liu2021} or cuprates \cite{Agterberg2020,cuprateReview}) and therefore take $\rho_k(T)\approx \text{const}$, $\Delta_0(T) = \Delta_0 \sqrt{1-T/T_c}$. As a result, the ratio is temperature independent. In contrast, for pure PDW, apart from direct vicinity of $T_{PDW}$ where finite-size effects are important, the ratio shows a strong, linear dependence on temperature.

For completeness, let us discuss the case where $T_{CDW}<T_c$, where a finite field critical current peak is expected to appear below $T_{CDW}$ and can not be distinguished from a PDW-induced peak. In that case, however, the ordinary superconductivity could be suppressed by application of a magnetic field, revealing the CDW nature of the state, whereas a PDW would be suppressed simultaneously with ordinary SC.

%Finally, we note that coexistence scenario between PDW and ordinary superconductivity does not present any obstacle for our approach, as the finite-field peak in critical current should only form below the PDW onset temperature and can not be conflated with 

%Lastly, we consider the coexistence of PDW and uniform superconductivity, which is thought to explain experimental results in a cuprate compound \cite{Agterberg2020,Edkins2019}. The superconducting phases of the PDW and uniform order parameter are locked within a layer \cite{Agterberg2020}, as a result of which our first harmonic term contains tunneling of both $q=0$ and finite-$\vec{q}$ Cooper pairs. Similar to the CDW + uniform case, where $T_{CDW} > T_{sc}$, the critical Josephson current exhibits a finite and zero field peak. In contrast to this case, once we go above $T_c$ for either the PDW or uniform order parameter, one of the peaks will vanish.  

{\it Discussion and outlook---} While we have focused on the case of single-$\vec{q}$ PDW states, generalization of our results is straightforward to multi-$\vec{q}$ PDWs. Each additional PDW modulation wave-vector in Eq. \eqref{eq:pdwOP}, will introduce an additional finite-field peak along a different orientation (seen as an additional bright spot in Fig. \ref{fig:vary_orientation}), as will follow from adding the contributions in Eq. \eqref{eq:I1}. While additional cross-terms between different PDWs can appear, at low twist angles they will oscillate on much shorter scales than the ones produced by the moir\'e patterns of the respective PDWs.

%allows us to define an additional moire length scale, $\lambda_m^i$ and maximize the critical current for a unique combination of field orientation and $\Phi_{max}$ similar to the single $\vec{q}$ case shown in Fig. \ref{fig:vary_orientation}. 

In conclusion, we have demonstrated a method of detecting and characterizing pair density wave superconductors at a macroscopic scale using twist Josephson junctions. In the presence of in-plane magnetic field, the critical Josephson current of PDW junctions exhibits a maximum at a finite field value, that serves as a direct signature of the PDW state. By analyzing the temperature dependence of the critical current, one can deduce finer effects of higher-order coupling between PDWs across the twisted interface and distinguish them from the cases where PDW arises from coexistence of other order parameters. Furthermore, certain scenarios of disordered and fluctuating PDW that may not produce consistent PDW signatures in the critical current, are demonstrated to show robust finite-field peaks in Josephson current fluctuations. In contrast to the presently used and proposed techniques to study PDWs, our proposal constitutes a transport measurement and may be realized already using present experimental techniques.

%s of phases that generate a PDW-like order and surface PDW states.  

%We have presented a phase-sensitive probe of the PDW phase and its wave-vector by studying the Josephson effect in the presence of an in-plane magnetic field. Our proposal can probe the PDW phase at at macroscopic scale and differentiate the PDW order from coexistences of phases that generate a PDW-like order and surface PDW states. We have demonstrated that our proposal alone is sufficient to determine the direction of the spatial modulation and its indifference to junction geometry. We also introduced experimental protocols in the presence of strong disorder and for long junctions.

\bibliography{ref}

\begin{thebibliography}{80}%
\makeatletter
\providecommand \@ifxundefined [1]{%
 \@ifx{#1\undefined}
}%
\providecommand \@ifnum [1]{%
 \ifnum #1\expandafter \@firstoftwo
 \else \expandafter \@secondoftwo
 \fi
}%
\providecommand \@ifx [1]{%
 \ifx #1\expandafter \@firstoftwo
 \else \expandafter \@secondoftwo
 \fi
}%
\providecommand \natexlab [1]{#1}%
\providecommand \enquote  [1]{``#1''}%
\providecommand \bibnamefont  [1]{#1}%
\providecommand \bibfnamefont [1]{#1}%
\providecommand \citenamefont [1]{#1}%
\providecommand \href@noop [0]{\@secondoftwo}%
\providecommand \href [0]{\begingroup \@sanitize@url \@href}%
\providecommand \@href[1]{\@@startlink{#1}\@@href}%
\providecommand \@@href[1]{\endgroup#1\@@endlink}%
\providecommand \@sanitize@url [0]{\catcode `\\12\catcode `\$12\catcode `\&12\catcode `\#12\catcode `\^12\catcode `\_12\catcode `\%12\relax}%
\providecommand \@@startlink[1]{}%
\providecommand \@@endlink[0]{}%
\providecommand \url  [0]{\begingroup\@sanitize@url \@url }%
\providecommand \@url [1]{\endgroup\@href {#1}{\urlprefix }}%
\providecommand \urlprefix  [0]{URL }%
\providecommand \Eprint [0]{\href }%
\providecommand \doibase [0]{https://doi.org/}%
\providecommand \selectlanguage [0]{\@gobble}%
\providecommand \bibinfo  [0]{\@secondoftwo}%
\providecommand \bibfield  [0]{\@secondoftwo}%
\providecommand \translation [1]{[#1]}%
\providecommand \BibitemOpen [0]{}%
\providecommand \bibitemStop [0]{}%
\providecommand \bibitemNoStop [0]{.\EOS\space}%
\providecommand \EOS [0]{\spacefactor3000\relax}%
\providecommand \BibitemShut  [1]{\csname bibitem#1\endcsname}%
\let\auto@bib@innerbib\@empty
%</preamble>
\bibitem [{\citenamefont {Agterberg}\ \emph {et~al.}(2020)\citenamefont {Agterberg}, \citenamefont {Davis}, \citenamefont {Edkins}, \citenamefont {Fradkin}, \citenamefont {Van~Harlingen}, \citenamefont {Kivelson}, \citenamefont {Lee}, \citenamefont {Radzihovsky}, \citenamefont {Tranquada},\ and\ \citenamefont {Wang}}]{Agterberg2020}%
  \BibitemOpen
  \bibfield  {author} {\bibinfo {author} {\bibfnamefont {D.~F.}\ \bibnamefont {Agterberg}}, \bibinfo {author} {\bibfnamefont {J.~S.}\ \bibnamefont {Davis}}, \bibinfo {author} {\bibfnamefont {S.~D.}\ \bibnamefont {Edkins}}, \bibinfo {author} {\bibfnamefont {E.}~\bibnamefont {Fradkin}}, \bibinfo {author} {\bibfnamefont {D.~J.}\ \bibnamefont {Van~Harlingen}}, \bibinfo {author} {\bibfnamefont {S.~A.}\ \bibnamefont {Kivelson}}, \bibinfo {author} {\bibfnamefont {P.~A.}\ \bibnamefont {Lee}}, \bibinfo {author} {\bibfnamefont {L.}~\bibnamefont {Radzihovsky}}, \bibinfo {author} {\bibfnamefont {J.~M.}\ \bibnamefont {Tranquada}},\ and\ \bibinfo {author} {\bibfnamefont {Y.}~\bibnamefont {Wang}},\ }\bibfield  {title} {\bibinfo {title} {The physics of pair-density waves: Cuprate superconductors and beyond},\ }\href {https://doi.org/https://doi.org/10.1146/annurev-conmatphys-031119-050711} {\bibfield  {journal} {\bibinfo  {journal} {Annual Review of Condensed Matter Physics}\ }\textbf {\bibinfo {volume} {11}},\ \bibinfo
  {pages} {231} (\bibinfo {year} {2020})}\BibitemShut {NoStop}%
\bibitem [{\citenamefont {Larkin}\ and\ \citenamefont {Ovchinnikov}(1964)}]{Larkin1964}%
  \BibitemOpen
  \bibfield  {author} {\bibinfo {author} {\bibfnamefont {A.~I.}\ \bibnamefont {Larkin}}\ and\ \bibinfo {author} {\bibfnamefont {Y.~N.}\ \bibnamefont {Ovchinnikov}},\ }\bibfield  {title} {\bibinfo {title} {{Nonuniform state of superconductors}},\ }\href@noop {} {\bibfield  {journal} {\bibinfo  {journal} {Zh. Eksp. Teor. Fiz.}\ }\textbf {\bibinfo {volume} {47}},\ \bibinfo {pages} {1136} (\bibinfo {year} {1964})}\BibitemShut {NoStop}%
\bibitem [{\citenamefont {Fulde}\ and\ \citenamefont {Ferrell}(1964)}]{Fulde1964}%
  \BibitemOpen
  \bibfield  {author} {\bibinfo {author} {\bibfnamefont {P.}~\bibnamefont {Fulde}}\ and\ \bibinfo {author} {\bibfnamefont {R.~A.}\ \bibnamefont {Ferrell}},\ }\bibfield  {title} {\bibinfo {title} {Superconductivity in a strong spin-exchange field},\ }\href {https://doi.org/10.1103/PhysRev.135.A550} {\bibfield  {journal} {\bibinfo  {journal} {Phys. Rev.}\ }\textbf {\bibinfo {volume} {135}},\ \bibinfo {pages} {A550} (\bibinfo {year} {1964})}\BibitemShut {NoStop}%
\bibitem [{\citenamefont {Liu}\ \emph {et~al.}(2024)\citenamefont {Liu}, \citenamefont {Huang}, \citenamefont {Huang}, \citenamefont {Moritz},\ and\ \citenamefont {Devereaux}}]{deveraux2024}%
  \BibitemOpen
  \bibfield  {author} {\bibinfo {author} {\bibfnamefont {F.}~\bibnamefont {Liu}}, \bibinfo {author} {\bibfnamefont {X.-X.}\ \bibnamefont {Huang}}, \bibinfo {author} {\bibfnamefont {E.~W.}\ \bibnamefont {Huang}}, \bibinfo {author} {\bibfnamefont {B.}~\bibnamefont {Moritz}},\ and\ \bibinfo {author} {\bibfnamefont {T.~P.}\ \bibnamefont {Devereaux}},\ }\bibfield  {title} {\bibinfo {title} {Enhanced pair-density-wave vertices in a bilayer hubbard model at half filling},\ }\href {https://doi.org/10.1103/PhysRevLett.133.156503} {\bibfield  {journal} {\bibinfo  {journal} {Phys. Rev. Lett.}\ }\textbf {\bibinfo {volume} {133}},\ \bibinfo {pages} {156503} (\bibinfo {year} {2024})}\BibitemShut {NoStop}%
\bibitem [{\citenamefont {Huang}\ \emph {et~al.}(2022)\citenamefont {Huang}, \citenamefont {Han}, \citenamefont {Kivelson},\ and\ \citenamefont {Yao}}]{Huang2022}%
  \BibitemOpen
  \bibfield  {author} {\bibinfo {author} {\bibfnamefont {K.~S.}\ \bibnamefont {Huang}}, \bibinfo {author} {\bibfnamefont {Z.}~\bibnamefont {Han}}, \bibinfo {author} {\bibfnamefont {S.~A.}\ \bibnamefont {Kivelson}},\ and\ \bibinfo {author} {\bibfnamefont {H.}~\bibnamefont {Yao}},\ }\bibfield  {title} {\bibinfo {title} {Pair-density-wave in the strong coupling limit of the holstein-hubbard model},\ }\bibfield  {journal} {\bibinfo  {journal} {npj Quantum Materials}\ }\textbf {\bibinfo {volume} {7}},\ \href {https://doi.org/10.1038/s41535-022-00426-w} {10.1038/s41535-022-00426-w} (\bibinfo {year} {2022})\BibitemShut {NoStop}%
\bibitem [{\citenamefont {Wang}\ \emph {et~al.}(2025{\natexlab{a}})\citenamefont {Wang}, \citenamefont {Sun}, \citenamefont {Wang}, \citenamefont {Han}, \citenamefont {Kivelson},\ and\ \citenamefont {Yao}}]{hongyao_2025}%
  \BibitemOpen
  \bibfield  {author} {\bibinfo {author} {\bibfnamefont {J.}~\bibnamefont {Wang}}, \bibinfo {author} {\bibfnamefont {W.}~\bibnamefont {Sun}}, \bibinfo {author} {\bibfnamefont {H.-X.}\ \bibnamefont {Wang}}, \bibinfo {author} {\bibfnamefont {Z.}~\bibnamefont {Han}}, \bibinfo {author} {\bibfnamefont {S.~A.}\ \bibnamefont {Kivelson}},\ and\ \bibinfo {author} {\bibfnamefont {H.}~\bibnamefont {Yao}},\ }\bibfield  {title} {\bibinfo {title} {Pair-density-wave phase of strongly interacting electrons on the triangular lattice: A variational monte carlo study},\ }\href {https://doi.org/10.1103/gvw1-xk98} {\bibfield  {journal} {\bibinfo  {journal} {Phys. Rev. B}\ }\textbf {\bibinfo {volume} {112}},\ \bibinfo {pages} {L140505} (\bibinfo {year} {2025}{\natexlab{a}})}\BibitemShut {NoStop}%
\bibitem [{\citenamefont {Himeda}\ \emph {et~al.}(2002)\citenamefont {Himeda}, \citenamefont {Kato},\ and\ \citenamefont {Ogata}}]{himeda2002}%
  \BibitemOpen
  \bibfield  {author} {\bibinfo {author} {\bibfnamefont {A.}~\bibnamefont {Himeda}}, \bibinfo {author} {\bibfnamefont {T.}~\bibnamefont {Kato}},\ and\ \bibinfo {author} {\bibfnamefont {M.}~\bibnamefont {Ogata}},\ }\bibfield  {title} {\bibinfo {title} {Stripe states with spatially oscillating $\mathit{d}$-wave superconductivity in the two-dimensional $\mathit{t}\ensuremath{-}{\mathit{t}}^{\ensuremath{'}}\ensuremath{-}\mathit{J}$ model},\ }\href {https://doi.org/10.1103/PhysRevLett.88.117001} {\bibfield  {journal} {\bibinfo  {journal} {Phys. Rev. Lett.}\ }\textbf {\bibinfo {volume} {88}},\ \bibinfo {pages} {117001} (\bibinfo {year} {2002})}\BibitemShut {NoStop}%
\bibitem [{\citenamefont {Berg}\ \emph {et~al.}(2007)\citenamefont {Berg}, \citenamefont {Fradkin}, \citenamefont {Kim}, \citenamefont {Kivelson}, \citenamefont {Oganesyan}, \citenamefont {Tranquada},\ and\ \citenamefont {Zhang}}]{Berg2007}%
  \BibitemOpen
  \bibfield  {author} {\bibinfo {author} {\bibfnamefont {E.}~\bibnamefont {Berg}}, \bibinfo {author} {\bibfnamefont {E.}~\bibnamefont {Fradkin}}, \bibinfo {author} {\bibfnamefont {E.-A.}\ \bibnamefont {Kim}}, \bibinfo {author} {\bibfnamefont {S.~A.}\ \bibnamefont {Kivelson}}, \bibinfo {author} {\bibfnamefont {V.}~\bibnamefont {Oganesyan}}, \bibinfo {author} {\bibfnamefont {J.~M.}\ \bibnamefont {Tranquada}},\ and\ \bibinfo {author} {\bibfnamefont {S.~C.}\ \bibnamefont {Zhang}},\ }\bibfield  {title} {\bibinfo {title} {Dynamical layer decoupling in a stripe-ordered high-${T}_{c}$ superconductor},\ }\href {https://doi.org/10.1103/PhysRevLett.99.127003} {\bibfield  {journal} {\bibinfo  {journal} {Phys. Rev. Lett.}\ }\textbf {\bibinfo {volume} {99}},\ \bibinfo {pages} {127003} (\bibinfo {year} {2007})}\BibitemShut {NoStop}%
\bibitem [{\citenamefont {Berg}\ \emph {et~al.}(2009{\natexlab{a}})\citenamefont {Berg}, \citenamefont {Fradkin}, \citenamefont {Kivelson},\ and\ \citenamefont {Tranquada}}]{Berg2009newphy}%
  \BibitemOpen
  \bibfield  {author} {\bibinfo {author} {\bibfnamefont {E.}~\bibnamefont {Berg}}, \bibinfo {author} {\bibfnamefont {E.}~\bibnamefont {Fradkin}}, \bibinfo {author} {\bibfnamefont {S.~A.}\ \bibnamefont {Kivelson}},\ and\ \bibinfo {author} {\bibfnamefont {J.~M.}\ \bibnamefont {Tranquada}},\ }\bibfield  {title} {\bibinfo {title} {Striped superconductors: how spin, charge and superconducting orders intertwine in the cuprates},\ }\href {https://doi.org/10.1088/1367-2630/11/11/115004} {\bibfield  {journal} {\bibinfo  {journal} {New Journal of Physics}\ }\textbf {\bibinfo {volume} {11}},\ \bibinfo {pages} {115004} (\bibinfo {year} {2009}{\natexlab{a}})}\BibitemShut {NoStop}%
\bibitem [{\citenamefont {Loder}\ \emph {et~al.}(2011)\citenamefont {Loder}, \citenamefont {Graser}, \citenamefont {Kampf},\ and\ \citenamefont {Kopp}}]{kopp_2011}%
  \BibitemOpen
  \bibfield  {author} {\bibinfo {author} {\bibfnamefont {F.}~\bibnamefont {Loder}}, \bibinfo {author} {\bibfnamefont {S.}~\bibnamefont {Graser}}, \bibinfo {author} {\bibfnamefont {A.~P.}\ \bibnamefont {Kampf}},\ and\ \bibinfo {author} {\bibfnamefont {T.}~\bibnamefont {Kopp}},\ }\bibfield  {title} {\bibinfo {title} {Mean-field pairing theory for the charge-stripe phase of high-temperature cuprate superconductors},\ }\href {https://doi.org/10.1103/PhysRevLett.107.187001} {\bibfield  {journal} {\bibinfo  {journal} {Phys. Rev. Lett.}\ }\textbf {\bibinfo {volume} {107}},\ \bibinfo {pages} {187001} (\bibinfo {year} {2011})}\BibitemShut {NoStop}%
\bibitem [{\citenamefont {Soto-Garrido}\ and\ \citenamefont {Fradkin}(2014)}]{sotogarrido2014}%
  \BibitemOpen
  \bibfield  {author} {\bibinfo {author} {\bibfnamefont {R.}~\bibnamefont {Soto-Garrido}}\ and\ \bibinfo {author} {\bibfnamefont {E.}~\bibnamefont {Fradkin}},\ }\bibfield  {title} {\bibinfo {title} {Pair-density-wave superconducting states and electronic liquid-crystal phases},\ }\href {https://doi.org/10.1103/PhysRevB.89.165126} {\bibfield  {journal} {\bibinfo  {journal} {Phys. Rev. B}\ }\textbf {\bibinfo {volume} {89}},\ \bibinfo {pages} {165126} (\bibinfo {year} {2014})}\BibitemShut {NoStop}%
\bibitem [{\citenamefont {Vafek}\ \emph {et~al.}(2014)\citenamefont {Vafek}, \citenamefont {Murray},\ and\ \citenamefont {Cvetkovic}}]{vafek2014}%
  \BibitemOpen
  \bibfield  {author} {\bibinfo {author} {\bibfnamefont {O.}~\bibnamefont {Vafek}}, \bibinfo {author} {\bibfnamefont {J.~M.}\ \bibnamefont {Murray}},\ and\ \bibinfo {author} {\bibfnamefont {V.}~\bibnamefont {Cvetkovic}},\ }\bibfield  {title} {\bibinfo {title} {Superconductivity on the brink of spin-charge order in a doped honeycomb bilayer},\ }\href {https://doi.org/10.1103/PhysRevLett.112.147002} {\bibfield  {journal} {\bibinfo  {journal} {Phys. Rev. Lett.}\ }\textbf {\bibinfo {volume} {112}},\ \bibinfo {pages} {147002} (\bibinfo {year} {2014})}\BibitemShut {NoStop}%
\bibitem [{\citenamefont {Wang}\ \emph {et~al.}(2015)\citenamefont {Wang}, \citenamefont {Agterberg},\ and\ \citenamefont {Chubukov}}]{chubukov2015}%
  \BibitemOpen
  \bibfield  {author} {\bibinfo {author} {\bibfnamefont {Y.}~\bibnamefont {Wang}}, \bibinfo {author} {\bibfnamefont {D.~F.}\ \bibnamefont {Agterberg}},\ and\ \bibinfo {author} {\bibfnamefont {A.}~\bibnamefont {Chubukov}},\ }\bibfield  {title} {\bibinfo {title} {Coexistence of charge-density-wave and pair-density-wave orders in underdoped cuprates},\ }\href {https://doi.org/10.1103/PhysRevLett.114.197001} {\bibfield  {journal} {\bibinfo  {journal} {Phys. Rev. Lett.}\ }\textbf {\bibinfo {volume} {114}},\ \bibinfo {pages} {197001} (\bibinfo {year} {2015})}\BibitemShut {NoStop}%
\bibitem [{\citenamefont {Fradkin}\ \emph {et~al.}(2015)\citenamefont {Fradkin}, \citenamefont {Kivelson},\ and\ \citenamefont {Tranquada}}]{intertwined_2015}%
  \BibitemOpen
  \bibfield  {author} {\bibinfo {author} {\bibfnamefont {E.}~\bibnamefont {Fradkin}}, \bibinfo {author} {\bibfnamefont {S.~A.}\ \bibnamefont {Kivelson}},\ and\ \bibinfo {author} {\bibfnamefont {J.~M.}\ \bibnamefont {Tranquada}},\ }\bibfield  {title} {\bibinfo {title} {Colloquium: Theory of intertwined orders in high temperature superconductors},\ }\href {https://doi.org/10.1103/RevModPhys.87.457} {\bibfield  {journal} {\bibinfo  {journal} {Rev. Mod. Phys.}\ }\textbf {\bibinfo {volume} {87}},\ \bibinfo {pages} {457} (\bibinfo {year} {2015})}\BibitemShut {NoStop}%
\bibitem [{\citenamefont {Lee}(2014)}]{Lee2014}%
  \BibitemOpen
  \bibfield  {author} {\bibinfo {author} {\bibfnamefont {P.~A.}\ \bibnamefont {Lee}},\ }\bibfield  {title} {\bibinfo {title} {Amperean pairing and the pseudogap phase of cuprate superconductors},\ }\href {https://doi.org/10.1103/PhysRevX.4.031017} {\bibfield  {journal} {\bibinfo  {journal} {Phys. Rev. X}\ }\textbf {\bibinfo {volume} {4}},\ \bibinfo {pages} {031017} (\bibinfo {year} {2014})}\BibitemShut {NoStop}%
\bibitem [{\citenamefont {Kargarian}\ \emph {et~al.}(2016)\citenamefont {Kargarian}, \citenamefont {Efimkin},\ and\ \citenamefont {Galitski}}]{kargarian_2016}%
  \BibitemOpen
  \bibfield  {author} {\bibinfo {author} {\bibfnamefont {M.}~\bibnamefont {Kargarian}}, \bibinfo {author} {\bibfnamefont {D.~K.}\ \bibnamefont {Efimkin}},\ and\ \bibinfo {author} {\bibfnamefont {V.}~\bibnamefont {Galitski}},\ }\bibfield  {title} {\bibinfo {title} {Amperean pairing at the surface of topological insulators},\ }\href {https://doi.org/10.1103/PhysRevLett.117.076806} {\bibfield  {journal} {\bibinfo  {journal} {Phys. Rev. Lett.}\ }\textbf {\bibinfo {volume} {117}},\ \bibinfo {pages} {076806} (\bibinfo {year} {2016})}\BibitemShut {NoStop}%
\bibitem [{\citenamefont {Wu}\ \emph {et~al.}(2023{\natexlab{a}})\citenamefont {Wu}, \citenamefont {Nosov}, \citenamefont {Patel},\ and\ \citenamefont {Raghu}}]{yiming_2023}%
  \BibitemOpen
  \bibfield  {author} {\bibinfo {author} {\bibfnamefont {Y.-M.}\ \bibnamefont {Wu}}, \bibinfo {author} {\bibfnamefont {P.~A.}\ \bibnamefont {Nosov}}, \bibinfo {author} {\bibfnamefont {A.~A.}\ \bibnamefont {Patel}},\ and\ \bibinfo {author} {\bibfnamefont {S.}~\bibnamefont {Raghu}},\ }\bibfield  {title} {\bibinfo {title} {Pair density wave order from electron repulsion},\ }\href {https://doi.org/10.1103/PhysRevLett.130.026001} {\bibfield  {journal} {\bibinfo  {journal} {Phys. Rev. Lett.}\ }\textbf {\bibinfo {volume} {130}},\ \bibinfo {pages} {026001} (\bibinfo {year} {2023}{\natexlab{a}})}\BibitemShut {NoStop}%
\bibitem [{\citenamefont {Setty}\ \emph {et~al.}(2023{\natexlab{a}})\citenamefont {Setty}, \citenamefont {Fanfarillo},\ and\ \citenamefont {Hirschfeld}}]{Setty2023}%
  \BibitemOpen
  \bibfield  {author} {\bibinfo {author} {\bibfnamefont {C.}~\bibnamefont {Setty}}, \bibinfo {author} {\bibfnamefont {L.}~\bibnamefont {Fanfarillo}},\ and\ \bibinfo {author} {\bibfnamefont {P.~J.}\ \bibnamefont {Hirschfeld}},\ }\bibfield  {title} {\bibinfo {title} {Mechanism for fluctuating pair density wave},\ }\bibfield  {journal} {\bibinfo  {journal} {Nature Communications}\ }\textbf {\bibinfo {volume} {14}},\ \href {https://doi.org/10.1038/s41467-023-38956-x} {10.1038/s41467-023-38956-x} (\bibinfo {year} {2023}{\natexlab{a}})\BibitemShut {NoStop}%
\bibitem [{\citenamefont {Setty}\ \emph {et~al.}(2023{\natexlab{b}})\citenamefont {Setty}, \citenamefont {Zhao}, \citenamefont {Fanfarillo}, \citenamefont {Huang}, \citenamefont {Hirschfeld}, \citenamefont {Phillips},\ and\ \citenamefont {Yang}}]{kunyang2023}%
  \BibitemOpen
  \bibfield  {author} {\bibinfo {author} {\bibfnamefont {C.}~\bibnamefont {Setty}}, \bibinfo {author} {\bibfnamefont {J.}~\bibnamefont {Zhao}}, \bibinfo {author} {\bibfnamefont {L.}~\bibnamefont {Fanfarillo}}, \bibinfo {author} {\bibfnamefont {E.~W.}\ \bibnamefont {Huang}}, \bibinfo {author} {\bibfnamefont {P.~J.}\ \bibnamefont {Hirschfeld}}, \bibinfo {author} {\bibfnamefont {P.~W.}\ \bibnamefont {Phillips}},\ and\ \bibinfo {author} {\bibfnamefont {K.}~\bibnamefont {Yang}},\ }\bibfield  {title} {\bibinfo {title} {Exact solution for finite center-of-mass momentum cooper pairing},\ }\href {https://doi.org/10.1103/PhysRevB.108.174506} {\bibfield  {journal} {\bibinfo  {journal} {Phys. Rev. B}\ }\textbf {\bibinfo {volume} {108}},\ \bibinfo {pages} {174506} (\bibinfo {year} {2023}{\natexlab{b}})}\BibitemShut {NoStop}%
\bibitem [{\citenamefont {Berg}\ \emph {et~al.}(2010)\citenamefont {Berg}, \citenamefont {Fradkin},\ and\ \citenamefont {Kivelson}}]{Berg2010}%
  \BibitemOpen
  \bibfield  {author} {\bibinfo {author} {\bibfnamefont {E.}~\bibnamefont {Berg}}, \bibinfo {author} {\bibfnamefont {E.}~\bibnamefont {Fradkin}},\ and\ \bibinfo {author} {\bibfnamefont {S.~A.}\ \bibnamefont {Kivelson}},\ }\bibfield  {title} {\bibinfo {title} {Pair-density-wave correlations in the kondo-heisenberg model},\ }\href {https://doi.org/10.1103/PhysRevLett.105.146403} {\bibfield  {journal} {\bibinfo  {journal} {Phys. Rev. Lett.}\ }\textbf {\bibinfo {volume} {105}},\ \bibinfo {pages} {146403} (\bibinfo {year} {2010})}\BibitemShut {NoStop}%
\bibitem [{\citenamefont {Coleman}\ \emph {et~al.}(2022)\citenamefont {Coleman}, \citenamefont {Panigrahi},\ and\ \citenamefont {Tsvelik}}]{cpt_2022}%
  \BibitemOpen
  \bibfield  {author} {\bibinfo {author} {\bibfnamefont {P.}~\bibnamefont {Coleman}}, \bibinfo {author} {\bibfnamefont {A.}~\bibnamefont {Panigrahi}},\ and\ \bibinfo {author} {\bibfnamefont {A.}~\bibnamefont {Tsvelik}},\ }\bibfield  {title} {\bibinfo {title} {Solvable 3d kondo lattice exhibiting pair density wave, odd-frequency pairing, and order fractionalization},\ }\href {https://doi.org/10.1103/PhysRevLett.129.177601} {\bibfield  {journal} {\bibinfo  {journal} {Phys. Rev. Lett.}\ }\textbf {\bibinfo {volume} {129}},\ \bibinfo {pages} {177601} (\bibinfo {year} {2022})}\BibitemShut {NoStop}%
\bibitem [{\citenamefont {Zhang}\ and\ \citenamefont {Vishwanath}(2022)}]{yahui2022}%
  \BibitemOpen
  \bibfield  {author} {\bibinfo {author} {\bibfnamefont {Y.-H.}\ \bibnamefont {Zhang}}\ and\ \bibinfo {author} {\bibfnamefont {A.}~\bibnamefont {Vishwanath}},\ }\bibfield  {title} {\bibinfo {title} {Pair-density-wave superconductor from doping haldane chain and rung-singlet ladder},\ }\href {https://doi.org/10.1103/PhysRevB.106.045103} {\bibfield  {journal} {\bibinfo  {journal} {Phys. Rev. B}\ }\textbf {\bibinfo {volume} {106}},\ \bibinfo {pages} {045103} (\bibinfo {year} {2022})}\BibitemShut {NoStop}%
\bibitem [{\citenamefont {Liu}\ and\ \citenamefont {Han}(2024)}]{zhaoyu2025}%
  \BibitemOpen
  \bibfield  {author} {\bibinfo {author} {\bibfnamefont {F.}~\bibnamefont {Liu}}\ and\ \bibinfo {author} {\bibfnamefont {Z.}~\bibnamefont {Han}},\ }\bibfield  {title} {\bibinfo {title} {{Pair density wave and $\mathit{s}\ifmmode\pm\else\textpm\fi{}\mathit{id}$ superconductivity in a strongly coupled lightly doped Kondo insulator}},\ }\href {https://doi.org/10.1103/PhysRevB.109.L121101} {\bibfield  {journal} {\bibinfo  {journal} {Phys. Rev. B}\ }\textbf {\bibinfo {volume} {109}},\ \bibinfo {pages} {L121101} (\bibinfo {year} {2024})}\BibitemShut {NoStop}%
\bibitem [{\citenamefont {Roy}\ and\ \citenamefont {Herbut}(2010)}]{bitan2010}%
  \BibitemOpen
  \bibfield  {author} {\bibinfo {author} {\bibfnamefont {B.}~\bibnamefont {Roy}}\ and\ \bibinfo {author} {\bibfnamefont {I.~F.}\ \bibnamefont {Herbut}},\ }\bibfield  {title} {\bibinfo {title} {Unconventional superconductivity on honeycomb lattice: Theory of kekule order parameter},\ }\href {https://doi.org/10.1103/PhysRevB.82.035429} {\bibfield  {journal} {\bibinfo  {journal} {Phys. Rev. B}\ }\textbf {\bibinfo {volume} {82}},\ \bibinfo {pages} {035429} (\bibinfo {year} {2010})}\BibitemShut {NoStop}%
\bibitem [{\citenamefont {Venderley}\ and\ \citenamefont {Kim}(2019)}]{Venderley2019}%
  \BibitemOpen
  \bibfield  {author} {\bibinfo {author} {\bibfnamefont {J.}~\bibnamefont {Venderley}}\ and\ \bibinfo {author} {\bibfnamefont {E.-A.}\ \bibnamefont {Kim}},\ }\bibfield  {title} {\bibinfo {title} {Evidence of pair-density wave in spin-valley locked systems},\ }\bibfield  {journal} {\bibinfo  {journal} {Science Advances}\ }\textbf {\bibinfo {volume} {5}},\ \href {https://doi.org/10.1126/sciadv.aat4698} {10.1126/sciadv.aat4698} (\bibinfo {year} {2019})\BibitemShut {NoStop}%
\bibitem [{\citenamefont {Han}\ and\ \citenamefont {Kivelson}(2022)}]{zhaoyu2022}%
  \BibitemOpen
  \bibfield  {author} {\bibinfo {author} {\bibfnamefont {Z.}~\bibnamefont {Han}}\ and\ \bibinfo {author} {\bibfnamefont {S.~A.}\ \bibnamefont {Kivelson}},\ }\bibfield  {title} {\bibinfo {title} {Pair density wave and reentrant superconducting tendencies originating from valley polarization},\ }\href {https://doi.org/10.1103/PhysRevB.105.L100509} {\bibfield  {journal} {\bibinfo  {journal} {Phys. Rev. B}\ }\textbf {\bibinfo {volume} {105}},\ \bibinfo {pages} {L100509} (\bibinfo {year} {2022})}\BibitemShut {NoStop}%
\bibitem [{\citenamefont {Shaffer}\ \emph {et~al.}(2023)\citenamefont {Shaffer}, \citenamefont {Burnell},\ and\ \citenamefont {Fernandes}}]{Shaffer2023}%
  \BibitemOpen
  \bibfield  {author} {\bibinfo {author} {\bibfnamefont {D.}~\bibnamefont {Shaffer}}, \bibinfo {author} {\bibfnamefont {F.~J.}\ \bibnamefont {Burnell}},\ and\ \bibinfo {author} {\bibfnamefont {R.~M.}\ \bibnamefont {Fernandes}},\ }\bibfield  {title} {\bibinfo {title} {Weak-coupling theory of pair density wave instabilities in transition metal dichalcogenides},\ }\href {https://doi.org/10.1103/PhysRevB.107.224516} {\bibfield  {journal} {\bibinfo  {journal} {Phys. Rev. B}\ }\textbf {\bibinfo {volume} {107}},\ \bibinfo {pages} {224516} (\bibinfo {year} {2023})}\BibitemShut {NoStop}%
\bibitem [{\citenamefont {Jiang}\ and\ \citenamefont {Yao}(2024)}]{jiang2024}%
  \BibitemOpen
  \bibfield  {author} {\bibinfo {author} {\bibfnamefont {Y.-F.}\ \bibnamefont {Jiang}}\ and\ \bibinfo {author} {\bibfnamefont {H.}~\bibnamefont {Yao}},\ }\bibfield  {title} {\bibinfo {title} {Pair-density-wave superconductivity: A microscopic model on the 2d honeycomb lattice},\ }\href {https://doi.org/10.1103/PhysRevLett.133.176501} {\bibfield  {journal} {\bibinfo  {journal} {Phys. Rev. Lett.}\ }\textbf {\bibinfo {volume} {133}},\ \bibinfo {pages} {176501} (\bibinfo {year} {2024})}\BibitemShut {NoStop}%
\bibitem [{\citenamefont {Chou}\ \emph {et~al.}(2025)\citenamefont {Chou}, \citenamefont {Zhu},\ and\ \citenamefont {Das~Sarma}}]{sarma2025}%
  \BibitemOpen
  \bibfield  {author} {\bibinfo {author} {\bibfnamefont {Y.-Z.}\ \bibnamefont {Chou}}, \bibinfo {author} {\bibfnamefont {J.}~\bibnamefont {Zhu}},\ and\ \bibinfo {author} {\bibfnamefont {S.}~\bibnamefont {Das~Sarma}},\ }\bibfield  {title} {\bibinfo {title} {Intravalley spin-polarized superconductivity in rhombohedral tetralayer graphene},\ }\href {https://doi.org/10.1103/PhysRevB.111.174523} {\bibfield  {journal} {\bibinfo  {journal} {Phys. Rev. B}\ }\textbf {\bibinfo {volume} {111}},\ \bibinfo {pages} {174523} (\bibinfo {year} {2025})}\BibitemShut {NoStop}%
\bibitem [{\citenamefont {Wu}\ \emph {et~al.}(2023{\natexlab{b}})\citenamefont {Wu}, \citenamefont {Wu},\ and\ \citenamefont {Wu}}]{fcwu2023}%
  \BibitemOpen
  \bibfield  {author} {\bibinfo {author} {\bibfnamefont {Z.}~\bibnamefont {Wu}}, \bibinfo {author} {\bibfnamefont {Y.-M.}\ \bibnamefont {Wu}},\ and\ \bibinfo {author} {\bibfnamefont {F.}~\bibnamefont {Wu}},\ }\bibfield  {title} {\bibinfo {title} {Pair density wave and loop current promoted by van hove singularities in moir\'e systems},\ }\href {https://doi.org/10.1103/PhysRevB.107.045122} {\bibfield  {journal} {\bibinfo  {journal} {Phys. Rev. B}\ }\textbf {\bibinfo {volume} {107}},\ \bibinfo {pages} {045122} (\bibinfo {year} {2023}{\natexlab{b}})}\BibitemShut {NoStop}%
\bibitem [{\citenamefont {Scammell}\ \emph {et~al.}(2023)\citenamefont {Scammell}, \citenamefont {Ingham}, \citenamefont {Li},\ and\ \citenamefont {Sushkov}}]{Scammell2023}%
  \BibitemOpen
  \bibfield  {author} {\bibinfo {author} {\bibfnamefont {H.~D.}\ \bibnamefont {Scammell}}, \bibinfo {author} {\bibfnamefont {J.}~\bibnamefont {Ingham}}, \bibinfo {author} {\bibfnamefont {T.}~\bibnamefont {Li}},\ and\ \bibinfo {author} {\bibfnamefont {O.~P.}\ \bibnamefont {Sushkov}},\ }\bibfield  {title} {\bibinfo {title} {Chiral excitonic order from twofold van hove singularities in kagome metals},\ }\href {https://doi.org/10.1038/s41467-023-35987-2} {\bibfield  {journal} {\bibinfo  {journal} {Nature Communications}\ }\textbf {\bibinfo {volume} {14}},\ \bibinfo {pages} {605} (\bibinfo {year} {2023})}\BibitemShut {NoStop}%
\bibitem [{\citenamefont {Schwemmer}\ \emph {et~al.}(2024)\citenamefont {Schwemmer}, \citenamefont {Hohmann}, \citenamefont {D\"urrnagel}, \citenamefont {Potten}, \citenamefont {Beyer}, \citenamefont {Rachel}, \citenamefont {Wu}, \citenamefont {Raghu}, \citenamefont {M\"uller}, \citenamefont {Hanke},\ and\ \citenamefont {Thomale}}]{Schwemmer2024}%
  \BibitemOpen
  \bibfield  {author} {\bibinfo {author} {\bibfnamefont {T.}~\bibnamefont {Schwemmer}}, \bibinfo {author} {\bibfnamefont {H.}~\bibnamefont {Hohmann}}, \bibinfo {author} {\bibfnamefont {M.}~\bibnamefont {D\"urrnagel}}, \bibinfo {author} {\bibfnamefont {J.}~\bibnamefont {Potten}}, \bibinfo {author} {\bibfnamefont {J.}~\bibnamefont {Beyer}}, \bibinfo {author} {\bibfnamefont {S.}~\bibnamefont {Rachel}}, \bibinfo {author} {\bibfnamefont {Y.-M.}\ \bibnamefont {Wu}}, \bibinfo {author} {\bibfnamefont {S.}~\bibnamefont {Raghu}}, \bibinfo {author} {\bibfnamefont {T.}~\bibnamefont {M\"uller}}, \bibinfo {author} {\bibfnamefont {W.}~\bibnamefont {Hanke}},\ and\ \bibinfo {author} {\bibfnamefont {R.}~\bibnamefont {Thomale}},\ }\bibfield  {title} {\bibinfo {title} {Sublattice modulated superconductivity in the kagome hubbard model},\ }\href {https://doi.org/10.1103/PhysRevB.110.024501} {\bibfield  {journal} {\bibinfo  {journal} {Phys. Rev. B}\ }\textbf {\bibinfo {volume} {110}},\ \bibinfo {pages} {024501} (\bibinfo {year}
  {2024})}\BibitemShut {NoStop}%
\bibitem [{\citenamefont {Wu}\ \emph {et~al.}(2023{\natexlab{c}})\citenamefont {Wu}, \citenamefont {Thomale},\ and\ \citenamefont {Raghu}}]{Wu2023}%
  \BibitemOpen
  \bibfield  {author} {\bibinfo {author} {\bibfnamefont {Y.-M.}\ \bibnamefont {Wu}}, \bibinfo {author} {\bibfnamefont {R.}~\bibnamefont {Thomale}},\ and\ \bibinfo {author} {\bibfnamefont {S.}~\bibnamefont {Raghu}},\ }\bibfield  {title} {\bibinfo {title} {Sublattice interference promotes pair density wave order in kagome metals},\ }\href {https://doi.org/10.1103/PhysRevB.108.L081117} {\bibfield  {journal} {\bibinfo  {journal} {Phys. Rev. B}\ }\textbf {\bibinfo {volume} {108}},\ \bibinfo {pages} {L081117} (\bibinfo {year} {2023}{\natexlab{c}})}\BibitemShut {NoStop}%
\bibitem [{\citenamefont {Yao}\ \emph {et~al.}(2025)\citenamefont {Yao}, \citenamefont {Wang}, \citenamefont {Wang}, \citenamefont {Yin},\ and\ \citenamefont {Wang}}]{Yao2025}%
  \BibitemOpen
  \bibfield  {author} {\bibinfo {author} {\bibfnamefont {M.}~\bibnamefont {Yao}}, \bibinfo {author} {\bibfnamefont {Y.}~\bibnamefont {Wang}}, \bibinfo {author} {\bibfnamefont {D.}~\bibnamefont {Wang}}, \bibinfo {author} {\bibfnamefont {J.-X.}\ \bibnamefont {Yin}},\ and\ \bibinfo {author} {\bibfnamefont {Q.-H.}\ \bibnamefont {Wang}},\ }\bibfield  {title} {\bibinfo {title} {Self-consistent theory of $2\ifmmode\times\else\texttimes\fi{}2$ pair density waves in kagome superconductors},\ }\href {https://doi.org/10.1103/PhysRevB.111.094505} {\bibfield  {journal} {\bibinfo  {journal} {Phys. Rev. B}\ }\textbf {\bibinfo {volume} {111}},\ \bibinfo {pages} {094505} (\bibinfo {year} {2025})}\BibitemShut {NoStop}%
\bibitem [{\citenamefont {Gil}\ and\ \citenamefont {Berg}(2025)}]{gil2025}%
  \BibitemOpen
  \bibfield  {author} {\bibinfo {author} {\bibfnamefont {A.}~\bibnamefont {Gil}}\ and\ \bibinfo {author} {\bibfnamefont {E.}~\bibnamefont {Berg}},\ }\href {https://arxiv.org/abs/2504.19321} {\bibinfo {title} {Charge and pair density waves in a spin and valley-polarized system at a van-hove singularity}} (\bibinfo {year} {2025}),\ \Eprint {https://arxiv.org/abs/2504.19321} {arXiv:2504.19321 [cond-mat.str-el]} \BibitemShut {NoStop}%
\bibitem [{\citenamefont {Cho}\ \emph {et~al.}(2012)\citenamefont {Cho}, \citenamefont {Bardarson}, \citenamefont {Lu},\ and\ \citenamefont {Moore}}]{cho_2012}%
  \BibitemOpen
  \bibfield  {author} {\bibinfo {author} {\bibfnamefont {G.~Y.}\ \bibnamefont {Cho}}, \bibinfo {author} {\bibfnamefont {J.~H.}\ \bibnamefont {Bardarson}}, \bibinfo {author} {\bibfnamefont {Y.-M.}\ \bibnamefont {Lu}},\ and\ \bibinfo {author} {\bibfnamefont {J.~E.}\ \bibnamefont {Moore}},\ }\bibfield  {title} {\bibinfo {title} {Superconductivity of doped weyl semimetals: Finite-momentum pairing and electronic analog of the ${}^{3}$he-$a$ phase},\ }\href {https://doi.org/10.1103/PhysRevB.86.214514} {\bibfield  {journal} {\bibinfo  {journal} {Phys. Rev. B}\ }\textbf {\bibinfo {volume} {86}},\ \bibinfo {pages} {214514} (\bibinfo {year} {2012})}\BibitemShut {NoStop}%
\bibitem [{\citenamefont {Zhou}\ and\ \citenamefont {Wang}(2022)}]{Zhou2022}%
  \BibitemOpen
  \bibfield  {author} {\bibinfo {author} {\bibfnamefont {S.}~\bibnamefont {Zhou}}\ and\ \bibinfo {author} {\bibfnamefont {Z.}~\bibnamefont {Wang}},\ }\bibfield  {title} {\bibinfo {title} {Chern fermi pocket, topological pair density wave, and charge-4e and charge-6e superconductivity in kagomé superconductors},\ }\bibfield  {journal} {\bibinfo  {journal} {Nature Communications}\ }\textbf {\bibinfo {volume} {13}},\ \href {https://doi.org/10.1038/s41467-022-34832-2} {10.1038/s41467-022-34832-2} (\bibinfo {year} {2022})\BibitemShut {NoStop}%
\bibitem [{\citenamefont {Han}\ \emph {et~al.}(2024)\citenamefont {Han}, \citenamefont {Herzog-Arbeitman}, \citenamefont {Bernevig},\ and\ \citenamefont {Kivelson}}]{zhaoyu2024}%
  \BibitemOpen
  \bibfield  {author} {\bibinfo {author} {\bibfnamefont {Z.}~\bibnamefont {Han}}, \bibinfo {author} {\bibfnamefont {J.}~\bibnamefont {Herzog-Arbeitman}}, \bibinfo {author} {\bibfnamefont {B.~A.}\ \bibnamefont {Bernevig}},\ and\ \bibinfo {author} {\bibfnamefont {S.~A.}\ \bibnamefont {Kivelson}},\ }\bibfield  {title} {\bibinfo {title} {``quantum geometric nesting'' and solvable model flat-band systems},\ }\href {https://doi.org/10.1103/PhysRevX.14.041004} {\bibfield  {journal} {\bibinfo  {journal} {Phys. Rev. X}\ }\textbf {\bibinfo {volume} {14}},\ \bibinfo {pages} {041004} (\bibinfo {year} {2024})}\BibitemShut {NoStop}%
\bibitem [{\citenamefont {Slagle}\ and\ \citenamefont {Fu}(2020)}]{slagle_2020}%
  \BibitemOpen
  \bibfield  {author} {\bibinfo {author} {\bibfnamefont {K.}~\bibnamefont {Slagle}}\ and\ \bibinfo {author} {\bibfnamefont {L.}~\bibnamefont {Fu}},\ }\bibfield  {title} {\bibinfo {title} {Charge transfer excitations, pair density waves, and superconductivity in moir\'e materials},\ }\href {https://doi.org/10.1103/PhysRevB.102.235423} {\bibfield  {journal} {\bibinfo  {journal} {Phys. Rev. B}\ }\textbf {\bibinfo {volume} {102}},\ \bibinfo {pages} {235423} (\bibinfo {year} {2020})}\BibitemShut {NoStop}%
\bibitem [{\citenamefont {Wu}\ \emph {et~al.}(2023{\natexlab{d}})\citenamefont {Wu}, \citenamefont {Wu},\ and\ \citenamefont {Yao}}]{wu2003_prl}%
  \BibitemOpen
  \bibfield  {author} {\bibinfo {author} {\bibfnamefont {Y.-M.}\ \bibnamefont {Wu}}, \bibinfo {author} {\bibfnamefont {Z.}~\bibnamefont {Wu}},\ and\ \bibinfo {author} {\bibfnamefont {H.}~\bibnamefont {Yao}},\ }\bibfield  {title} {\bibinfo {title} {Pair-density-wave and chiral superconductivity in twisted bilayer transition metal dichalcogenides},\ }\href {https://doi.org/10.1103/PhysRevLett.130.126001} {\bibfield  {journal} {\bibinfo  {journal} {Phys. Rev. Lett.}\ }\textbf {\bibinfo {volume} {130}},\ \bibinfo {pages} {126001} (\bibinfo {year} {2023}{\natexlab{d}})}\BibitemShut {NoStop}%
\bibitem [{\citenamefont {Chen}\ and\ \citenamefont {Sheng}(2023)}]{sheng_2023}%
  \BibitemOpen
  \bibfield  {author} {\bibinfo {author} {\bibfnamefont {F.}~\bibnamefont {Chen}}\ and\ \bibinfo {author} {\bibfnamefont {D.~N.}\ \bibnamefont {Sheng}},\ }\bibfield  {title} {\bibinfo {title} {Singlet, triplet, and pair density wave superconductivity in the doped triangular-lattice moir\'e system},\ }\href {https://doi.org/10.1103/PhysRevB.108.L201110} {\bibfield  {journal} {\bibinfo  {journal} {Phys. Rev. B}\ }\textbf {\bibinfo {volume} {108}},\ \bibinfo {pages} {L201110} (\bibinfo {year} {2023})}\BibitemShut {NoStop}%
\bibitem [{\citenamefont {Wang}\ \emph {et~al.}(2025{\natexlab{b}})\citenamefont {Wang}, \citenamefont {Hu}, \citenamefont {Huang},\ and\ \citenamefont {Yao}}]{wang2025negativeroutepairdensity}%
  \BibitemOpen
  \bibfield  {author} {\bibinfo {author} {\bibfnamefont {H.-X.}\ \bibnamefont {Wang}}, \bibinfo {author} {\bibfnamefont {Y.-J.}\ \bibnamefont {Hu}}, \bibinfo {author} {\bibfnamefont {W.}~\bibnamefont {Huang}},\ and\ \bibinfo {author} {\bibfnamefont {H.}~\bibnamefont {Yao}},\ }\href {https://arxiv.org/abs/2512.06100} {\bibinfo {title} {A "negative" route to pair density wave order}} (\bibinfo {year} {2025}{\natexlab{b}}),\ \Eprint {https://arxiv.org/abs/2512.06100} {arXiv:2512.06100 [cond-mat.supr-con]} \BibitemShut {NoStop}%
\bibitem [{\citenamefont {Zhu}\ \emph {et~al.}(2025)\citenamefont {Zhu}, \citenamefont {Sun}, \citenamefont {Gong}, \citenamefont {Huang}, \citenamefont {Feng}, \citenamefont {Scalettar},\ and\ \citenamefont {Guo}}]{scalettar2025}%
  \BibitemOpen
  \bibfield  {author} {\bibinfo {author} {\bibfnamefont {X.}~\bibnamefont {Zhu}}, \bibinfo {author} {\bibfnamefont {J.}~\bibnamefont {Sun}}, \bibinfo {author} {\bibfnamefont {S.-S.}\ \bibnamefont {Gong}}, \bibinfo {author} {\bibfnamefont {W.}~\bibnamefont {Huang}}, \bibinfo {author} {\bibfnamefont {S.}~\bibnamefont {Feng}}, \bibinfo {author} {\bibfnamefont {R.~T.}\ \bibnamefont {Scalettar}},\ and\ \bibinfo {author} {\bibfnamefont {H.}~\bibnamefont {Guo}},\ }\bibfield  {title} {\bibinfo {title} {{Rigorous demonstration of pair-density-wave superconductivity in the ${\ensuremath{\sigma}}_{z}$-Hubbard model}},\ }\href {https://doi.org/10.1103/PhysRevB.111.045158} {\bibfield  {journal} {\bibinfo  {journal} {Phys. Rev. B}\ }\textbf {\bibinfo {volume} {111}},\ \bibinfo {pages} {045158} (\bibinfo {year} {2025})}\BibitemShut {NoStop}%
\bibitem [{\citenamefont {Hamidian}\ \emph {et~al.}(2016)\citenamefont {Hamidian}, \citenamefont {Edkins}, \citenamefont {Joo}, \citenamefont {Kostin}, \citenamefont {Eisaki}, \citenamefont {Uchida}, \citenamefont {Lawler}, \citenamefont {Kim}, \citenamefont {Mackenzie}, \citenamefont {Fujita}, \citenamefont {Lee},\ and\ \citenamefont {Davis}}]{Hamidian2016}%
  \BibitemOpen
  \bibfield  {author} {\bibinfo {author} {\bibfnamefont {M.~H.}\ \bibnamefont {Hamidian}}, \bibinfo {author} {\bibfnamefont {S.~D.}\ \bibnamefont {Edkins}}, \bibinfo {author} {\bibfnamefont {S.~H.}\ \bibnamefont {Joo}}, \bibinfo {author} {\bibfnamefont {A.}~\bibnamefont {Kostin}}, \bibinfo {author} {\bibfnamefont {H.}~\bibnamefont {Eisaki}}, \bibinfo {author} {\bibfnamefont {S.}~\bibnamefont {Uchida}}, \bibinfo {author} {\bibfnamefont {M.~J.}\ \bibnamefont {Lawler}}, \bibinfo {author} {\bibfnamefont {E.-A.}\ \bibnamefont {Kim}}, \bibinfo {author} {\bibfnamefont {A.~P.}\ \bibnamefont {Mackenzie}}, \bibinfo {author} {\bibfnamefont {K.}~\bibnamefont {Fujita}}, \bibinfo {author} {\bibfnamefont {J.}~\bibnamefont {Lee}},\ and\ \bibinfo {author} {\bibfnamefont {J.~C.~S.}\ \bibnamefont {Davis}},\ }\bibfield  {title} {\bibinfo {title} {Detection of a cooper-pair density wave in bi2sr2cacu2o8+x},\ }\href {https://doi.org/10.1038/nature17411} {\bibfield  {journal} {\bibinfo  {journal} {Nature}\ }\textbf {\bibinfo
  {volume} {532}},\ \bibinfo {pages} {343–347} (\bibinfo {year} {2016})}\BibitemShut {NoStop}%
\bibitem [{\citenamefont {Du}\ \emph {et~al.}(2020)\citenamefont {Du}, \citenamefont {Li}, \citenamefont {Joo}, \citenamefont {Donoway}, \citenamefont {Lee}, \citenamefont {Davis}, \citenamefont {Gu}, \citenamefont {Johnson},\ and\ \citenamefont {Fujita}}]{Du2020}%
  \BibitemOpen
  \bibfield  {author} {\bibinfo {author} {\bibfnamefont {Z.}~\bibnamefont {Du}}, \bibinfo {author} {\bibfnamefont {H.}~\bibnamefont {Li}}, \bibinfo {author} {\bibfnamefont {S.~H.}\ \bibnamefont {Joo}}, \bibinfo {author} {\bibfnamefont {E.~P.}\ \bibnamefont {Donoway}}, \bibinfo {author} {\bibfnamefont {J.}~\bibnamefont {Lee}}, \bibinfo {author} {\bibfnamefont {J.~C.~S.}\ \bibnamefont {Davis}}, \bibinfo {author} {\bibfnamefont {G.}~\bibnamefont {Gu}}, \bibinfo {author} {\bibfnamefont {P.~D.}\ \bibnamefont {Johnson}},\ and\ \bibinfo {author} {\bibfnamefont {K.}~\bibnamefont {Fujita}},\ }\bibfield  {title} {\bibinfo {title} {Imaging the energy gap modulations of the cuprate pair-density-wave state},\ }\href {https://doi.org/10.1038/s41586-020-2143-x} {\bibfield  {journal} {\bibinfo  {journal} {Nature}\ }\textbf {\bibinfo {volume} {580}},\ \bibinfo {pages} {65–70} (\bibinfo {year} {2020})}\BibitemShut {NoStop}%
\bibitem [{\citenamefont {Park}\ \emph {et~al.}(2012)\citenamefont {Park}, \citenamefont {Lee}, \citenamefont {Martin}, \citenamefont {Lu}, \citenamefont {Sidorov}, \citenamefont {Gofryk}, \citenamefont {Ronning}, \citenamefont {Bauer},\ and\ \citenamefont {Thompson}}]{hf1}%
  \BibitemOpen
  \bibfield  {author} {\bibinfo {author} {\bibfnamefont {T.}~\bibnamefont {Park}}, \bibinfo {author} {\bibfnamefont {H.}~\bibnamefont {Lee}}, \bibinfo {author} {\bibfnamefont {I.}~\bibnamefont {Martin}}, \bibinfo {author} {\bibfnamefont {X.}~\bibnamefont {Lu}}, \bibinfo {author} {\bibfnamefont {V.~A.}\ \bibnamefont {Sidorov}}, \bibinfo {author} {\bibfnamefont {K.}~\bibnamefont {Gofryk}}, \bibinfo {author} {\bibfnamefont {F.}~\bibnamefont {Ronning}}, \bibinfo {author} {\bibfnamefont {E.~D.}\ \bibnamefont {Bauer}},\ and\ \bibinfo {author} {\bibfnamefont {J.~D.}\ \bibnamefont {Thompson}},\ }\bibfield  {title} {\bibinfo {title} {Textured superconducting phase in the heavy fermion ${\mathrm{cerhin}}_{5}$},\ }\href {https://doi.org/10.1103/PhysRevLett.108.077003} {\bibfield  {journal} {\bibinfo  {journal} {Phys. Rev. Lett.}\ }\textbf {\bibinfo {volume} {108}},\ \bibinfo {pages} {077003} (\bibinfo {year} {2012})}\BibitemShut {NoStop}%
\bibitem [{\citenamefont {Gerber}\ \emph {et~al.}(2013)\citenamefont {Gerber}, \citenamefont {Bartkowiak}, \citenamefont {Gavilano}, \citenamefont {Ressouche}, \citenamefont {Egetenmeyer}, \citenamefont {Niedermayer}, \citenamefont {Bianchi}, \citenamefont {Movshovich}, \citenamefont {Bauer}, \citenamefont {Thompson},\ and\ \citenamefont {Kenzelmann}}]{hf2}%
  \BibitemOpen
  \bibfield  {author} {\bibinfo {author} {\bibfnamefont {S.}~\bibnamefont {Gerber}}, \bibinfo {author} {\bibfnamefont {M.}~\bibnamefont {Bartkowiak}}, \bibinfo {author} {\bibfnamefont {J.~L.}\ \bibnamefont {Gavilano}}, \bibinfo {author} {\bibfnamefont {E.}~\bibnamefont {Ressouche}}, \bibinfo {author} {\bibfnamefont {N.}~\bibnamefont {Egetenmeyer}}, \bibinfo {author} {\bibfnamefont {C.}~\bibnamefont {Niedermayer}}, \bibinfo {author} {\bibfnamefont {A.~D.}\ \bibnamefont {Bianchi}}, \bibinfo {author} {\bibfnamefont {R.}~\bibnamefont {Movshovich}}, \bibinfo {author} {\bibfnamefont {E.~D.}\ \bibnamefont {Bauer}}, \bibinfo {author} {\bibfnamefont {J.~D.}\ \bibnamefont {Thompson}},\ and\ \bibinfo {author} {\bibfnamefont {M.}~\bibnamefont {Kenzelmann}},\ }\bibfield  {title} {\bibinfo {title} {Switching of magnetic domains reveals spatially inhomogeneous superconductivity},\ }\href {https://doi.org/10.1038/nphys2833} {\bibfield  {journal} {\bibinfo  {journal} {Nature Physics}\ }\textbf {\bibinfo {volume} {10}},\
  \bibinfo {pages} {126–129} (\bibinfo {year} {2013})}\BibitemShut {NoStop}%
\bibitem [{\citenamefont {Zhao}\ \emph {et~al.}(2023)\citenamefont {Zhao}, \citenamefont {Blackwell}, \citenamefont {Thinel}, \citenamefont {Handa}, \citenamefont {Ishida}, \citenamefont {Zhu}, \citenamefont {Iyo}, \citenamefont {Eisaki}, \citenamefont {Pasupathy},\ and\ \citenamefont {Fujita}}]{fujita2023}%
  \BibitemOpen
  \bibfield  {author} {\bibinfo {author} {\bibfnamefont {H.}~\bibnamefont {Zhao}}, \bibinfo {author} {\bibfnamefont {R.}~\bibnamefont {Blackwell}}, \bibinfo {author} {\bibfnamefont {M.}~\bibnamefont {Thinel}}, \bibinfo {author} {\bibfnamefont {T.}~\bibnamefont {Handa}}, \bibinfo {author} {\bibfnamefont {S.}~\bibnamefont {Ishida}}, \bibinfo {author} {\bibfnamefont {X.}~\bibnamefont {Zhu}}, \bibinfo {author} {\bibfnamefont {A.}~\bibnamefont {Iyo}}, \bibinfo {author} {\bibfnamefont {H.}~\bibnamefont {Eisaki}}, \bibinfo {author} {\bibfnamefont {A.~N.}\ \bibnamefont {Pasupathy}},\ and\ \bibinfo {author} {\bibfnamefont {K.}~\bibnamefont {Fujita}},\ }\bibfield  {title} {\bibinfo {title} {Smectic pair-density-wave order in eurbfe4as4},\ }\href {https://doi.org/10.1038/s41586-023-06103-7} {\bibfield  {journal} {\bibinfo  {journal} {Nature}\ }\textbf {\bibinfo {volume} {618}},\ \bibinfo {pages} {940–945} (\bibinfo {year} {2023})}\BibitemShut {NoStop}%
\bibitem [{\citenamefont {Liu}\ \emph {et~al.}(2023)\citenamefont {Liu}, \citenamefont {Wei}, \citenamefont {He}, \citenamefont {Zhang}, \citenamefont {Wang},\ and\ \citenamefont {Wang}}]{Liu2023}%
  \BibitemOpen
  \bibfield  {author} {\bibinfo {author} {\bibfnamefont {Y.}~\bibnamefont {Liu}}, \bibinfo {author} {\bibfnamefont {T.}~\bibnamefont {Wei}}, \bibinfo {author} {\bibfnamefont {G.}~\bibnamefont {He}}, \bibinfo {author} {\bibfnamefont {Y.}~\bibnamefont {Zhang}}, \bibinfo {author} {\bibfnamefont {Z.}~\bibnamefont {Wang}},\ and\ \bibinfo {author} {\bibfnamefont {J.}~\bibnamefont {Wang}},\ }\bibfield  {title} {\bibinfo {title} {Pair density wave state in a monolayer high-tc iron-based superconductor},\ }\href {https://doi.org/10.1038/s41586-023-06072-x} {\bibfield  {journal} {\bibinfo  {journal} {Nature}\ }\textbf {\bibinfo {volume} {618}},\ \bibinfo {pages} {934} (\bibinfo {year} {2023})}\BibitemShut {NoStop}%
\bibitem [{\citenamefont {Gu}\ \emph {et~al.}(2023)\citenamefont {Gu}, \citenamefont {Carroll}, \citenamefont {Wang}, \citenamefont {Ran}, \citenamefont {Broyles}, \citenamefont {Siddiquee}, \citenamefont {Butch}, \citenamefont {Saha}, \citenamefont {Paglione}, \citenamefont {Davis},\ and\ \citenamefont {Liu}}]{Gu2023}%
  \BibitemOpen
  \bibfield  {author} {\bibinfo {author} {\bibfnamefont {Q.}~\bibnamefont {Gu}}, \bibinfo {author} {\bibfnamefont {J.~P.}\ \bibnamefont {Carroll}}, \bibinfo {author} {\bibfnamefont {S.}~\bibnamefont {Wang}}, \bibinfo {author} {\bibfnamefont {S.}~\bibnamefont {Ran}}, \bibinfo {author} {\bibfnamefont {C.}~\bibnamefont {Broyles}}, \bibinfo {author} {\bibfnamefont {H.}~\bibnamefont {Siddiquee}}, \bibinfo {author} {\bibfnamefont {N.~P.}\ \bibnamefont {Butch}}, \bibinfo {author} {\bibfnamefont {S.~R.}\ \bibnamefont {Saha}}, \bibinfo {author} {\bibfnamefont {J.}~\bibnamefont {Paglione}}, \bibinfo {author} {\bibfnamefont {J.~C.~S.}\ \bibnamefont {Davis}},\ and\ \bibinfo {author} {\bibfnamefont {X.}~\bibnamefont {Liu}},\ }\bibfield  {title} {\bibinfo {title} {Detection of a pair density wave state in ute2},\ }\href {https://doi.org/10.1038/s41586-023-05919-7} {\bibfield  {journal} {\bibinfo  {journal} {Nature}\ }\textbf {\bibinfo {volume} {618}},\ \bibinfo {pages} {921} (\bibinfo {year} {2023})}\BibitemShut {NoStop}%
\bibitem [{\citenamefont {Aishwarya}\ \emph {et~al.}(2023)\citenamefont {Aishwarya}, \citenamefont {May-Mann}, \citenamefont {Raghavan}, \citenamefont {Nie}, \citenamefont {Romanelli}, \citenamefont {Ran}, \citenamefont {Saha}, \citenamefont {Paglione}, \citenamefont {Butch}, \citenamefont {Fradkin},\ and\ \citenamefont {Madhavan}}]{Aishwarya2023}%
  \BibitemOpen
  \bibfield  {author} {\bibinfo {author} {\bibfnamefont {A.}~\bibnamefont {Aishwarya}}, \bibinfo {author} {\bibfnamefont {J.}~\bibnamefont {May-Mann}}, \bibinfo {author} {\bibfnamefont {A.}~\bibnamefont {Raghavan}}, \bibinfo {author} {\bibfnamefont {L.}~\bibnamefont {Nie}}, \bibinfo {author} {\bibfnamefont {M.}~\bibnamefont {Romanelli}}, \bibinfo {author} {\bibfnamefont {S.}~\bibnamefont {Ran}}, \bibinfo {author} {\bibfnamefont {S.~R.}\ \bibnamefont {Saha}}, \bibinfo {author} {\bibfnamefont {J.}~\bibnamefont {Paglione}}, \bibinfo {author} {\bibfnamefont {N.~P.}\ \bibnamefont {Butch}}, \bibinfo {author} {\bibfnamefont {E.}~\bibnamefont {Fradkin}},\ and\ \bibinfo {author} {\bibfnamefont {V.}~\bibnamefont {Madhavan}},\ }\bibfield  {title} {\bibinfo {title} {Magnetic-field-sensitive charge density waves in the superconductor ute2},\ }\href {https://doi.org/10.1038/s41586-023-06005-8} {\bibfield  {journal} {\bibinfo  {journal} {Nature}\ }\textbf {\bibinfo {volume} {618}},\ \bibinfo {pages} {928} (\bibinfo {year}
  {2023})}\BibitemShut {NoStop}%
\bibitem [{\citenamefont {Aishwarya}\ \emph {et~al.}(2024)\citenamefont {Aishwarya}, \citenamefont {May-Mann}, \citenamefont {Almoalem}, \citenamefont {Ran}, \citenamefont {Saha}, \citenamefont {Paglione}, \citenamefont {Butch}, \citenamefont {Fradkin},\ and\ \citenamefont {Madhavan}}]{Aishwarya2024}%
  \BibitemOpen
  \bibfield  {author} {\bibinfo {author} {\bibfnamefont {A.}~\bibnamefont {Aishwarya}}, \bibinfo {author} {\bibfnamefont {J.}~\bibnamefont {May-Mann}}, \bibinfo {author} {\bibfnamefont {A.}~\bibnamefont {Almoalem}}, \bibinfo {author} {\bibfnamefont {S.}~\bibnamefont {Ran}}, \bibinfo {author} {\bibfnamefont {S.~R.}\ \bibnamefont {Saha}}, \bibinfo {author} {\bibfnamefont {J.}~\bibnamefont {Paglione}}, \bibinfo {author} {\bibfnamefont {N.~P.}\ \bibnamefont {Butch}}, \bibinfo {author} {\bibfnamefont {E.}~\bibnamefont {Fradkin}},\ and\ \bibinfo {author} {\bibfnamefont {V.}~\bibnamefont {Madhavan}},\ }\bibfield  {title} {\bibinfo {title} {Melting of the charge density wave by generation of pairs of topological defects in ute2},\ }\href {https://doi.org/10.1038/s41567-024-02429-9} {\bibfield  {journal} {\bibinfo  {journal} {Nature Physics}\ }\textbf {\bibinfo {volume} {20}},\ \bibinfo {pages} {964} (\bibinfo {year} {2024})}\BibitemShut {NoStop}%
\bibitem [{\citenamefont {Chen}\ \emph {et~al.}(2021)\citenamefont {Chen}, \citenamefont {Yang}, \citenamefont {Hu}, \citenamefont {Zhao}, \citenamefont {Yuan}, \citenamefont {Xing}, \citenamefont {Qian}, \citenamefont {Huang}, \citenamefont {Li}, \citenamefont {Ye}, \citenamefont {Ma}, \citenamefont {Ni}, \citenamefont {Zhang}, \citenamefont {Yin}, \citenamefont {Gong}, \citenamefont {Tu}, \citenamefont {Lei}, \citenamefont {Tan}, \citenamefont {Zhou}, \citenamefont {Shen}, \citenamefont {Dong}, \citenamefont {Yan}, \citenamefont {Wang},\ and\ \citenamefont {Gao}}]{Chen2021}%
  \BibitemOpen
  \bibfield  {author} {\bibinfo {author} {\bibfnamefont {H.}~\bibnamefont {Chen}}, \bibinfo {author} {\bibfnamefont {H.}~\bibnamefont {Yang}}, \bibinfo {author} {\bibfnamefont {B.}~\bibnamefont {Hu}}, \bibinfo {author} {\bibfnamefont {Z.}~\bibnamefont {Zhao}}, \bibinfo {author} {\bibfnamefont {J.}~\bibnamefont {Yuan}}, \bibinfo {author} {\bibfnamefont {Y.}~\bibnamefont {Xing}}, \bibinfo {author} {\bibfnamefont {G.}~\bibnamefont {Qian}}, \bibinfo {author} {\bibfnamefont {Z.}~\bibnamefont {Huang}}, \bibinfo {author} {\bibfnamefont {G.}~\bibnamefont {Li}}, \bibinfo {author} {\bibfnamefont {Y.}~\bibnamefont {Ye}}, \bibinfo {author} {\bibfnamefont {S.}~\bibnamefont {Ma}}, \bibinfo {author} {\bibfnamefont {S.}~\bibnamefont {Ni}}, \bibinfo {author} {\bibfnamefont {H.}~\bibnamefont {Zhang}}, \bibinfo {author} {\bibfnamefont {Q.}~\bibnamefont {Yin}}, \bibinfo {author} {\bibfnamefont {C.}~\bibnamefont {Gong}}, \bibinfo {author} {\bibfnamefont {Z.}~\bibnamefont {Tu}}, \bibinfo {author} {\bibfnamefont {H.}~\bibnamefont
  {Lei}}, \bibinfo {author} {\bibfnamefont {H.}~\bibnamefont {Tan}}, \bibinfo {author} {\bibfnamefont {S.}~\bibnamefont {Zhou}}, \bibinfo {author} {\bibfnamefont {C.}~\bibnamefont {Shen}}, \bibinfo {author} {\bibfnamefont {X.}~\bibnamefont {Dong}}, \bibinfo {author} {\bibfnamefont {B.}~\bibnamefont {Yan}}, \bibinfo {author} {\bibfnamefont {Z.}~\bibnamefont {Wang}},\ and\ \bibinfo {author} {\bibfnamefont {H.-J.}\ \bibnamefont {Gao}},\ }\bibfield  {title} {\bibinfo {title} {Roton pair density wave in a strong-coupling kagome superconductor},\ }\href {https://doi.org/10.1038/s41586-021-03983-5} {\bibfield  {journal} {\bibinfo  {journal} {Nature}\ }\textbf {\bibinfo {volume} {599}},\ \bibinfo {pages} {222} (\bibinfo {year} {2021})}\BibitemShut {NoStop}%
\bibitem [{\citenamefont {Deng}\ \emph {et~al.}(2024)\citenamefont {Deng}, \citenamefont {Qin}, \citenamefont {Liu}, \citenamefont {Yang}, \citenamefont {Fu}, \citenamefont {Zhang}, \citenamefont {Wu}, \citenamefont {Wang}, \citenamefont {Shi}, \citenamefont {Liu}, \citenamefont {Liu}, \citenamefont {Yan}, \citenamefont {Song}, \citenamefont {Xu}, \citenamefont {Zhao}, \citenamefont {Yi}, \citenamefont {Xu}, \citenamefont {Hohmann}, \citenamefont {Holb{\ae}k}, \citenamefont {D{\"u}rrnagel}, \citenamefont {Zhou}, \citenamefont {Chang}, \citenamefont {Yao}, \citenamefont {Wang}, \citenamefont {Guguchia}, \citenamefont {Neupert}, \citenamefont {Thomale}, \citenamefont {Fischer},\ and\ \citenamefont {Yin}}]{Deng2024}%
  \BibitemOpen
  \bibfield  {author} {\bibinfo {author} {\bibfnamefont {H.}~\bibnamefont {Deng}}, \bibinfo {author} {\bibfnamefont {H.}~\bibnamefont {Qin}}, \bibinfo {author} {\bibfnamefont {G.}~\bibnamefont {Liu}}, \bibinfo {author} {\bibfnamefont {T.}~\bibnamefont {Yang}}, \bibinfo {author} {\bibfnamefont {R.}~\bibnamefont {Fu}}, \bibinfo {author} {\bibfnamefont {Z.}~\bibnamefont {Zhang}}, \bibinfo {author} {\bibfnamefont {X.}~\bibnamefont {Wu}}, \bibinfo {author} {\bibfnamefont {Z.}~\bibnamefont {Wang}}, \bibinfo {author} {\bibfnamefont {Y.}~\bibnamefont {Shi}}, \bibinfo {author} {\bibfnamefont {J.}~\bibnamefont {Liu}}, \bibinfo {author} {\bibfnamefont {H.}~\bibnamefont {Liu}}, \bibinfo {author} {\bibfnamefont {X.-Y.}\ \bibnamefont {Yan}}, \bibinfo {author} {\bibfnamefont {W.}~\bibnamefont {Song}}, \bibinfo {author} {\bibfnamefont {X.}~\bibnamefont {Xu}}, \bibinfo {author} {\bibfnamefont {Y.}~\bibnamefont {Zhao}}, \bibinfo {author} {\bibfnamefont {M.}~\bibnamefont {Yi}}, \bibinfo {author} {\bibfnamefont {G.}~\bibnamefont
  {Xu}}, \bibinfo {author} {\bibfnamefont {H.}~\bibnamefont {Hohmann}}, \bibinfo {author} {\bibfnamefont {S.~C.}\ \bibnamefont {Holb{\ae}k}}, \bibinfo {author} {\bibfnamefont {M.}~\bibnamefont {D{\"u}rrnagel}}, \bibinfo {author} {\bibfnamefont {S.}~\bibnamefont {Zhou}}, \bibinfo {author} {\bibfnamefont {G.}~\bibnamefont {Chang}}, \bibinfo {author} {\bibfnamefont {Y.}~\bibnamefont {Yao}}, \bibinfo {author} {\bibfnamefont {Q.}~\bibnamefont {Wang}}, \bibinfo {author} {\bibfnamefont {Z.}~\bibnamefont {Guguchia}}, \bibinfo {author} {\bibfnamefont {T.}~\bibnamefont {Neupert}}, \bibinfo {author} {\bibfnamefont {R.}~\bibnamefont {Thomale}}, \bibinfo {author} {\bibfnamefont {M.~H.}\ \bibnamefont {Fischer}},\ and\ \bibinfo {author} {\bibfnamefont {J.-X.}\ \bibnamefont {Yin}},\ }\bibfield  {title} {\bibinfo {title} {Chiral kagome superconductivity modulations with residual fermi arcs},\ }\href {https://doi.org/10.1038/s41586-024-07798-y} {\bibfield  {journal} {\bibinfo  {journal} {Nature}\ }\textbf {\bibinfo {volume}
  {632}},\ \bibinfo {pages} {775} (\bibinfo {year} {2024})}\BibitemShut {NoStop}%
\bibitem [{\citenamefont {Liu}\ \emph {et~al.}(2021)\citenamefont {Liu}, \citenamefont {Chong}, \citenamefont {Sharma},\ and\ \citenamefont {Davis}}]{Liu2021}%
  \BibitemOpen
  \bibfield  {author} {\bibinfo {author} {\bibfnamefont {X.}~\bibnamefont {Liu}}, \bibinfo {author} {\bibfnamefont {Y.~X.}\ \bibnamefont {Chong}}, \bibinfo {author} {\bibfnamefont {R.}~\bibnamefont {Sharma}},\ and\ \bibinfo {author} {\bibfnamefont {J.~C.~S.}\ \bibnamefont {Davis}},\ }\bibfield  {title} {\bibinfo {title} {Discovery of a cooper-pair density wave state in a transition-metal dichalcogenide},\ }\href {https://doi.org/10.1126/science.abd4607} {\bibfield  {journal} {\bibinfo  {journal} {Science}\ }\textbf {\bibinfo {volume} {372}},\ \bibinfo {pages} {1447} (\bibinfo {year} {2021})},\ \Eprint {https://arxiv.org/abs/https://www.science.org/doi/pdf/10.1126/science.abd4607} {https://www.science.org/doi/pdf/10.1126/science.abd4607} \BibitemShut {NoStop}%
\bibitem [{\citenamefont {Devarakonda}\ \emph {et~al.}(2024)\citenamefont {Devarakonda}, \citenamefont {Chen}, \citenamefont {Fang}, \citenamefont {Graf}, \citenamefont {Kriener}, \citenamefont {Akey}, \citenamefont {Bell}, \citenamefont {Suzuki},\ and\ \citenamefont {Checkelsky}}]{Devarakonda2024}%
  \BibitemOpen
  \bibfield  {author} {\bibinfo {author} {\bibfnamefont {A.}~\bibnamefont {Devarakonda}}, \bibinfo {author} {\bibfnamefont {A.}~\bibnamefont {Chen}}, \bibinfo {author} {\bibfnamefont {S.}~\bibnamefont {Fang}}, \bibinfo {author} {\bibfnamefont {D.}~\bibnamefont {Graf}}, \bibinfo {author} {\bibfnamefont {M.}~\bibnamefont {Kriener}}, \bibinfo {author} {\bibfnamefont {A.~J.}\ \bibnamefont {Akey}}, \bibinfo {author} {\bibfnamefont {D.~C.}\ \bibnamefont {Bell}}, \bibinfo {author} {\bibfnamefont {T.}~\bibnamefont {Suzuki}},\ and\ \bibinfo {author} {\bibfnamefont {J.~G.}\ \bibnamefont {Checkelsky}},\ }\bibfield  {title} {\bibinfo {title} {Evidence of striped electronic phases in a structurally modulated superlattice},\ }\href {https://doi.org/10.1038/s41586-024-07589-5} {\bibfield  {journal} {\bibinfo  {journal} {Nature}\ }\textbf {\bibinfo {volume} {631}},\ \bibinfo {pages} {526} (\bibinfo {year} {2024})}\BibitemShut {NoStop}%
\bibitem [{\citenamefont {Cao}\ \emph {et~al.}(2024)\citenamefont {Cao}, \citenamefont {Xue}, \citenamefont {Wang}, \citenamefont {Zhang}, \citenamefont {Kang}, \citenamefont {Gao}, \citenamefont {Mao},\ and\ \citenamefont {Jiang}}]{Cao2024}%
  \BibitemOpen
  \bibfield  {author} {\bibinfo {author} {\bibfnamefont {L.}~\bibnamefont {Cao}}, \bibinfo {author} {\bibfnamefont {Y.}~\bibnamefont {Xue}}, \bibinfo {author} {\bibfnamefont {Y.}~\bibnamefont {Wang}}, \bibinfo {author} {\bibfnamefont {F.-C.}\ \bibnamefont {Zhang}}, \bibinfo {author} {\bibfnamefont {J.}~\bibnamefont {Kang}}, \bibinfo {author} {\bibfnamefont {H.-J.}\ \bibnamefont {Gao}}, \bibinfo {author} {\bibfnamefont {J.}~\bibnamefont {Mao}},\ and\ \bibinfo {author} {\bibfnamefont {Y.}~\bibnamefont {Jiang}},\ }\bibfield  {title} {\bibinfo {title} {Directly visualizing nematic superconductivity driven by the pair density wave in nbse2},\ }\href {https://doi.org/10.1038/s41467-024-51558-5} {\bibfield  {journal} {\bibinfo  {journal} {Nature Communications}\ }\textbf {\bibinfo {volume} {15}},\ \bibinfo {pages} {7234} (\bibinfo {year} {2024})}\BibitemShut {NoStop}%
\bibitem [{\citenamefont {Mahmood}\ \emph {et~al.}(2022)\citenamefont {Mahmood}, \citenamefont {Devereaux}, \citenamefont {Abbamonte},\ and\ \citenamefont {Morr}}]{2earpes}%
  \BibitemOpen
  \bibfield  {author} {\bibinfo {author} {\bibfnamefont {F.}~\bibnamefont {Mahmood}}, \bibinfo {author} {\bibfnamefont {T.}~\bibnamefont {Devereaux}}, \bibinfo {author} {\bibfnamefont {P.}~\bibnamefont {Abbamonte}},\ and\ \bibinfo {author} {\bibfnamefont {D.~K.}\ \bibnamefont {Morr}},\ }\bibfield  {title} {\bibinfo {title} {Distinguishing finite-momentum superconducting pairing states with two-electron photoemission spectroscopy},\ }\href {https://doi.org/10.1103/PhysRevB.105.064515} {\bibfield  {journal} {\bibinfo  {journal} {Phys. Rev. B}\ }\textbf {\bibinfo {volume} {105}},\ \bibinfo {pages} {064515} (\bibinfo {year} {2022})}\BibitemShut {NoStop}%
\bibitem [{\citenamefont {Wu}\ \emph {et~al.}(2025)\citenamefont {Wu}, \citenamefont {Chubukov}, \citenamefont {Wang},\ and\ \citenamefont {Kivelson}}]{Wu2025}%
  \BibitemOpen
  \bibfield  {author} {\bibinfo {author} {\bibfnamefont {Y.-M.}\ \bibnamefont {Wu}}, \bibinfo {author} {\bibfnamefont {A.~V.}\ \bibnamefont {Chubukov}}, \bibinfo {author} {\bibfnamefont {Y.}~\bibnamefont {Wang}},\ and\ \bibinfo {author} {\bibfnamefont {S.~A.}\ \bibnamefont {Kivelson}},\ }\bibfield  {title} {\bibinfo {title} {Time-reversal symmetry breaking, collective modes, and raman spectrum in pair-density-wave states},\ }\href {https://doi.org/10.1038/s41535-025-00808-w} {\bibfield  {journal} {\bibinfo  {journal} {npj Quantum Materials}\ }\textbf {\bibinfo {volume} {10}},\ \bibinfo {pages} {84} (\bibinfo {year} {2025})}\BibitemShut {NoStop}%
\bibitem [{\citenamefont {Pixley}\ and\ \citenamefont {Volkov}(2025)}]{pixley2025}%
  \BibitemOpen
  \bibfield  {author} {\bibinfo {author} {\bibfnamefont {J.~H.}\ \bibnamefont {Pixley}}\ and\ \bibinfo {author} {\bibfnamefont {P.~A.}\ \bibnamefont {Volkov}},\ }\href {https://arxiv.org/abs/2503.23683} {\bibinfo {title} {Twisted nodal superconductors}} (\bibinfo {year} {2025}),\ \Eprint {https://arxiv.org/abs/2503.23683} {arXiv:2503.23683 [cond-mat.supr-con]} \BibitemShut {NoStop}%
\bibitem [{\citenamefont {Klemm*}(2005)}]{klemm2005phase}%
  \BibitemOpen
  \bibfield  {author} {\bibinfo {author} {\bibfnamefont {R.~A.}\ \bibnamefont {Klemm*}},\ }\bibfield  {title} {\bibinfo {title} {The phase-sensitive c-axis twist experiments on bi2sr2cacu2o8+ $\delta$ and their implications},\ }\href@noop {} {\bibfield  {journal} {\bibinfo  {journal} {Philosophical Magazine}\ }\textbf {\bibinfo {volume} {85}},\ \bibinfo {pages} {801} (\bibinfo {year} {2005})}\BibitemShut {NoStop}%
\bibitem [{\citenamefont {Xiao}\ \emph {et~al.}(2023)\citenamefont {Xiao}, \citenamefont {Vituri},\ and\ \citenamefont {Berg}}]{xiaoberg_2023}%
  \BibitemOpen
  \bibfield  {author} {\bibinfo {author} {\bibfnamefont {J.}~\bibnamefont {Xiao}}, \bibinfo {author} {\bibfnamefont {Y.}~\bibnamefont {Vituri}},\ and\ \bibinfo {author} {\bibfnamefont {E.}~\bibnamefont {Berg}},\ }\bibfield  {title} {\bibinfo {title} {Probing the order parameter symmetry of two-dimensional superconductors by twisted josephson interferometry},\ }\href {https://doi.org/10.1103/PhysRevB.108.094520} {\bibfield  {journal} {\bibinfo  {journal} {Phys. Rev. B}\ }\textbf {\bibinfo {volume} {108}},\ \bibinfo {pages} {094520} (\bibinfo {year} {2023})}\BibitemShut {NoStop}%
\bibitem [{\citenamefont {Yuan}\ and\ \citenamefont {Kivelson}(2024)}]{yuan2024phase}%
  \BibitemOpen
  \bibfield  {author} {\bibinfo {author} {\bibfnamefont {A.~C.}\ \bibnamefont {Yuan}}\ and\ \bibinfo {author} {\bibfnamefont {S.~A.}\ \bibnamefont {Kivelson}},\ }\bibfield  {title} {\bibinfo {title} {Phase sensitive information from a planar josephson junction},\ }\href@noop {} {\bibfield  {journal} {\bibinfo  {journal} {npj Quantum Materials}\ }\textbf {\bibinfo {volume} {9}},\ \bibinfo {pages} {93} (\bibinfo {year} {2024})}\BibitemShut {NoStop}%
\bibitem [{\citenamefont {Barone}\ and\ \citenamefont {Paterno}(1982)}]{barone1982}%
  \BibitemOpen
  \bibfield  {author} {\bibinfo {author} {\bibfnamefont {A.}~\bibnamefont {Barone}}\ and\ \bibinfo {author} {\bibfnamefont {G.}~\bibnamefont {Paterno}},\ }\href {https://books.google.com/books?id=FrjvAAAAMAAJ} {\emph {\bibinfo {title} {Physics and Applications of the Josephson Effect}}},\ A Wiley-interscience publication\ (\bibinfo  {publisher} {Wiley},\ \bibinfo {year} {1982})\BibitemShut {NoStop}%
\bibitem [{\citenamefont {Tummuru}\ \emph {et~al.}(2022)\citenamefont {Tummuru}, \citenamefont {Plugge},\ and\ \citenamefont {Franz}}]{Tummuru-Franz-PRB-2022}%
  \BibitemOpen
  \bibfield  {author} {\bibinfo {author} {\bibfnamefont {T.}~\bibnamefont {Tummuru}}, \bibinfo {author} {\bibfnamefont {S.}~\bibnamefont {Plugge}},\ and\ \bibinfo {author} {\bibfnamefont {M.}~\bibnamefont {Franz}},\ }\bibfield  {title} {\bibinfo {title} {Josephson effects in twisted cuprate bilayers},\ }\href {https://doi.org/10.1103/PhysRevB.105.064501} {\bibfield  {journal} {\bibinfo  {journal} {Phys. Rev. B}\ }\textbf {\bibinfo {volume} {105}},\ \bibinfo {pages} {064501} (\bibinfo {year} {2022})}\BibitemShut {NoStop}%
\bibitem [{\citenamefont {Volkov}\ \emph {et~al.}(2025)\citenamefont {Volkov}, \citenamefont {Zhao}, \citenamefont {Poccia}, \citenamefont {Cui}, \citenamefont {Kim},\ and\ \citenamefont {Pixley}}]{volkov_2025}%
  \BibitemOpen
  \bibfield  {author} {\bibinfo {author} {\bibfnamefont {P.~A.}\ \bibnamefont {Volkov}}, \bibinfo {author} {\bibfnamefont {S.~Y.~F.}\ \bibnamefont {Zhao}}, \bibinfo {author} {\bibfnamefont {N.}~\bibnamefont {Poccia}}, \bibinfo {author} {\bibfnamefont {X.}~\bibnamefont {Cui}}, \bibinfo {author} {\bibfnamefont {P.}~\bibnamefont {Kim}},\ and\ \bibinfo {author} {\bibfnamefont {J.~H.}\ \bibnamefont {Pixley}},\ }\bibfield  {title} {\bibinfo {title} {Josephson effects in twisted nodal superconductors},\ }\href {https://doi.org/10.1103/PhysRevB.111.014514} {\bibfield  {journal} {\bibinfo  {journal} {Phys. Rev. B}\ }\textbf {\bibinfo {volume} {111}},\ \bibinfo {pages} {014514} (\bibinfo {year} {2025})}\BibitemShut {NoStop}%
\bibitem [{\citenamefont {Agterberg}\ and\ \citenamefont {Tsunetsugu}(2008)}]{Agterberg2008}%
  \BibitemOpen
  \bibfield  {author} {\bibinfo {author} {\bibfnamefont {D.~F.}\ \bibnamefont {Agterberg}}\ and\ \bibinfo {author} {\bibfnamefont {H.}~\bibnamefont {Tsunetsugu}},\ }\bibfield  {title} {\bibinfo {title} {Dislocations and vortices in pair-density-wave superconductors},\ }\href {https://doi.org/10.1038/nphys999} {\bibfield  {journal} {\bibinfo  {journal} {Nature Physics}\ }\textbf {\bibinfo {volume} {4}},\ \bibinfo {pages} {639} (\bibinfo {year} {2008})}\BibitemShut {NoStop}%
\bibitem [{\citenamefont {Berg}\ \emph {et~al.}(2009{\natexlab{b}})\citenamefont {Berg}, \citenamefont {Fradkin},\ and\ \citenamefont {Kivelson}}]{Berg2009PRB}%
  \BibitemOpen
  \bibfield  {author} {\bibinfo {author} {\bibfnamefont {E.}~\bibnamefont {Berg}}, \bibinfo {author} {\bibfnamefont {E.}~\bibnamefont {Fradkin}},\ and\ \bibinfo {author} {\bibfnamefont {S.~A.}\ \bibnamefont {Kivelson}},\ }\bibfield  {title} {\bibinfo {title} {Theory of the striped superconductor},\ }\href {https://doi.org/10.1103/PhysRevB.79.064515} {\bibfield  {journal} {\bibinfo  {journal} {Phys. Rev. B}\ }\textbf {\bibinfo {volume} {79}},\ \bibinfo {pages} {064515} (\bibinfo {year} {2009}{\natexlab{b}})}\BibitemShut {NoStop}%
\bibitem [{\citenamefont {Can}\ \emph {et~al.}(2021)\citenamefont {Can}, \citenamefont {Tummuru}, \citenamefont {Day}, \citenamefont {Elfimov}, \citenamefont {Damascelli},\ and\ \citenamefont {Franz}}]{can2021high}%
  \BibitemOpen
  \bibfield  {author} {\bibinfo {author} {\bibfnamefont {O.}~\bibnamefont {Can}}, \bibinfo {author} {\bibfnamefont {T.}~\bibnamefont {Tummuru}}, \bibinfo {author} {\bibfnamefont {R.~P.}\ \bibnamefont {Day}}, \bibinfo {author} {\bibfnamefont {I.}~\bibnamefont {Elfimov}}, \bibinfo {author} {\bibfnamefont {A.}~\bibnamefont {Damascelli}},\ and\ \bibinfo {author} {\bibfnamefont {M.}~\bibnamefont {Franz}},\ }\bibfield  {title} {\bibinfo {title} {High-temperature topological superconductivity in twisted double-layer copper oxides},\ }\href@noop {} {\bibfield  {journal} {\bibinfo  {journal} {Nature Physics}\ }\textbf {\bibinfo {volume} {17}},\ \bibinfo {pages} {519} (\bibinfo {year} {2021})}\BibitemShut {NoStop}%
\bibitem [{\citenamefont {Volkov}\ \emph {et~al.}(2024)\citenamefont {Volkov}, \citenamefont {Lantagne-Hurtubise}, \citenamefont {Tummuru}, \citenamefont {Plugge}, \citenamefont {Pixley},\ and\ \citenamefont {Franz}}]{volkov_diode}%
  \BibitemOpen
  \bibfield  {author} {\bibinfo {author} {\bibfnamefont {P.~A.}\ \bibnamefont {Volkov}}, \bibinfo {author} {\bibfnamefont {E.}~\bibnamefont {Lantagne-Hurtubise}}, \bibinfo {author} {\bibfnamefont {T.}~\bibnamefont {Tummuru}}, \bibinfo {author} {\bibfnamefont {S.}~\bibnamefont {Plugge}}, \bibinfo {author} {\bibfnamefont {J.~H.}\ \bibnamefont {Pixley}},\ and\ \bibinfo {author} {\bibfnamefont {M.}~\bibnamefont {Franz}},\ }\bibfield  {title} {\bibinfo {title} {Josephson diode effects in twisted nodal superconductors},\ }\href {https://doi.org/10.1103/PhysRevB.109.094518} {\bibfield  {journal} {\bibinfo  {journal} {Phys. Rev. B}\ }\textbf {\bibinfo {volume} {109}},\ \bibinfo {pages} {094518} (\bibinfo {year} {2024})}\BibitemShut {NoStop}%
\bibitem [{\citenamefont {Tang}\ and\ \citenamefont {Volkov}(2025)}]{tang2025}%
  \BibitemOpen
  \bibfield  {author} {\bibinfo {author} {\bibfnamefont {J.}~\bibnamefont {Tang}}\ and\ \bibinfo {author} {\bibfnamefont {P.~A.}\ \bibnamefont {Volkov}},\ }\href {https://arxiv.org/abs/2507.22495} {\bibinfo {title} {Dynamical signatures and control of time-reversal breaking in twisted nodal superconductors}} (\bibinfo {year} {2025}),\ \Eprint {https://arxiv.org/abs/2507.22495} {arXiv:2507.22495 [cond-mat.supr-con]} \BibitemShut {NoStop}%
\bibitem [{\citenamefont {Golubov}\ \emph {et~al.}(2004)\citenamefont {Golubov}, \citenamefont {Kupriyanov},\ and\ \citenamefont {Il'ichev}}]{KupriyanovRev}%
  \BibitemOpen
  \bibfield  {author} {\bibinfo {author} {\bibfnamefont {A.~A.}\ \bibnamefont {Golubov}}, \bibinfo {author} {\bibfnamefont {M.~Y.}\ \bibnamefont {Kupriyanov}},\ and\ \bibinfo {author} {\bibfnamefont {E.}~\bibnamefont {Il'ichev}},\ }\bibfield  {title} {\bibinfo {title} {The current-phase relation in josephson junctions},\ }\href {https://doi.org/10.1103/RevModPhys.76.411} {\bibfield  {journal} {\bibinfo  {journal} {Rev. Mod. Phys.}\ }\textbf {\bibinfo {volume} {76}},\ \bibinfo {pages} {411} (\bibinfo {year} {2004})}\BibitemShut {NoStop}%
\bibitem [{Note1()}]{Note1}%
  \BibitemOpen
  \bibinfo {note} {For twist angles larger than $90^\circ $, the role of $q_+$ and $q_-$ as defined above, interchanges.}\BibitemShut {Stop}%
\bibitem [{\citenamefont {Owen}\ and\ \citenamefont {Scalapino}(1967)}]{owen1967}%
  \BibitemOpen
  \bibfield  {author} {\bibinfo {author} {\bibfnamefont {C.~S.}\ \bibnamefont {Owen}}\ and\ \bibinfo {author} {\bibfnamefont {D.~J.}\ \bibnamefont {Scalapino}},\ }\bibfield  {title} {\bibinfo {title} {Vortex structure and critical currents in josephson junctions},\ }\href {https://doi.org/10.1103/PhysRev.164.538} {\bibfield  {journal} {\bibinfo  {journal} {Phys. Rev.}\ }\textbf {\bibinfo {volume} {164}},\ \bibinfo {pages} {538} (\bibinfo {year} {1967})}\BibitemShut {NoStop}%
\bibitem [{Note2()}]{Note2}%
  \BibitemOpen
  \bibinfo {note} {Which depends on material parameters, such as the London penetration depth, but grows with decreasing Josephson coupling, i.e. weak junctions are can be usually considered short.}\BibitemShut {Stop}%
\bibitem [{\citenamefont {Tinkham}(2004)}]{tinkham2004}%
  \BibitemOpen
  \bibfield  {author} {\bibinfo {author} {\bibfnamefont {M.}~\bibnamefont {Tinkham}},\ }\href {https://books.google.com/books?id=VpUk3NfwDIkC} {\emph {\bibinfo {title} {Introduction to Superconductivity}}},\ Dover Books on Physics Series\ (\bibinfo  {publisher} {Dover Publications},\ \bibinfo {year} {2004})\BibitemShut {NoStop}%
\bibitem [{\citenamefont {Ambegaokar}\ and\ \citenamefont {Baratoff}(1963)}]{amb_bar_1963}%
  \BibitemOpen
  \bibfield  {author} {\bibinfo {author} {\bibfnamefont {V.}~\bibnamefont {Ambegaokar}}\ and\ \bibinfo {author} {\bibfnamefont {A.}~\bibnamefont {Baratoff}},\ }\bibfield  {title} {\bibinfo {title} {Tunneling between superconductors},\ }\href {https://doi.org/10.1103/PhysRevLett.10.486} {\bibfield  {journal} {\bibinfo  {journal} {Phys. Rev. Lett.}\ }\textbf {\bibinfo {volume} {10}},\ \bibinfo {pages} {486} (\bibinfo {year} {1963})}\BibitemShut {NoStop}%
\bibitem [{\citenamefont {Chen}\ \emph {et~al.}(2022)\citenamefont {Chen}, \citenamefont {Ren}, \citenamefont {Kennedy}, \citenamefont {Hamidian}, \citenamefont {Uchida}, \citenamefont {Eisaki}, \citenamefont {Johnson}, \citenamefont {O’Mahony},\ and\ \citenamefont {Davis}}]{Chen2022}%
  \BibitemOpen
  \bibfield  {author} {\bibinfo {author} {\bibfnamefont {W.}~\bibnamefont {Chen}}, \bibinfo {author} {\bibfnamefont {W.}~\bibnamefont {Ren}}, \bibinfo {author} {\bibfnamefont {N.}~\bibnamefont {Kennedy}}, \bibinfo {author} {\bibfnamefont {M.~H.}\ \bibnamefont {Hamidian}}, \bibinfo {author} {\bibfnamefont {S.}~\bibnamefont {Uchida}}, \bibinfo {author} {\bibfnamefont {H.}~\bibnamefont {Eisaki}}, \bibinfo {author} {\bibfnamefont {P.~D.}\ \bibnamefont {Johnson}}, \bibinfo {author} {\bibfnamefont {S.~M.}\ \bibnamefont {O’Mahony}},\ and\ \bibinfo {author} {\bibfnamefont {J.~C.~S.}\ \bibnamefont {Davis}},\ }\bibfield  {title} {\bibinfo {title} {Identification of a nematic pair density wave state in bi 2 sr 2 cacu 2 o 8+x},\ }\bibfield  {journal} {\bibinfo  {journal} {Proceedings of the National Academy of Sciences}\ }\textbf {\bibinfo {volume} {119}},\ \href {https://doi.org/10.1073/pnas.2206481119} {10.1073/pnas.2206481119} (\bibinfo {year} {2022})\BibitemShut {NoStop}%
\bibitem [{\citenamefont {Wang}\ \emph {et~al.}(2025{\natexlab{c}})\citenamefont {Wang}, \citenamefont {Xu}, \citenamefont {Li}, \citenamefont {Fan}, \citenamefont {Yang},\ and\ \citenamefont {Wen}}]{wen2025}%
  \BibitemOpen
  \bibfield  {author} {\bibinfo {author} {\bibfnamefont {Z.}~\bibnamefont {Wang}}, \bibinfo {author} {\bibfnamefont {J.}~\bibnamefont {Xu}}, \bibinfo {author} {\bibfnamefont {H.}~\bibnamefont {Li}}, \bibinfo {author} {\bibfnamefont {S.}~\bibnamefont {Fan}}, \bibinfo {author} {\bibfnamefont {H.}~\bibnamefont {Yang}},\ and\ \bibinfo {author} {\bibfnamefont {H.-H.}\ \bibnamefont {Wen}},\ }\bibfield  {title} {\bibinfo {title} {Evidence of pair density waves pinned by zn impurities in ${\mathrm{bi}}_{2}{\mathrm{sr}}_{2}{\mathrm{ca}(\mathrm{cu}}_{1\ensuremath{-}x}{\mathrm{zn}}_{x}{)}_{2}{\mathrm{o}}_{8+\ensuremath{\delta}}$},\ }\href {https://doi.org/10.1103/1jyq-vkls} {\bibfield  {journal} {\bibinfo  {journal} {Phys. Rev. B}\ }\textbf {\bibinfo {volume} {112}},\ \bibinfo {pages} {064502} (\bibinfo {year} {2025}{\natexlab{c}})}\BibitemShut {NoStop}%
\bibitem [{\citenamefont {Keimer}\ \emph {et~al.}(2015)\citenamefont {Keimer}, \citenamefont {Kivelson}, \citenamefont {Norman}, \citenamefont {Uchida},\ and\ \citenamefont {Zaanen}}]{cuprateReview}%
  \BibitemOpen
  \bibfield  {author} {\bibinfo {author} {\bibfnamefont {B.}~\bibnamefont {Keimer}}, \bibinfo {author} {\bibfnamefont {S.~A.}\ \bibnamefont {Kivelson}}, \bibinfo {author} {\bibfnamefont {M.~R.}\ \bibnamefont {Norman}}, \bibinfo {author} {\bibfnamefont {S.}~\bibnamefont {Uchida}},\ and\ \bibinfo {author} {\bibfnamefont {J.}~\bibnamefont {Zaanen}},\ }\bibfield  {title} {\bibinfo {title} {From quantum matter to high-temperature superconductivity in copper oxides},\ }\href {https://doi.org/10.1038/nature14165} {\bibfield  {journal} {\bibinfo  {journal} {Nature}\ }\textbf {\bibinfo {volume} {518}},\ \bibinfo {pages} {179} (\bibinfo {year} {2015})}\BibitemShut {NoStop}%
\end{thebibliography}%

\newpage
\clearpage
\onecolumngrid
\appendix

\renewcommand{\thefigure}{S\arabic{figure}}
\addtocounter{equation}{-11}
\addtocounter{figure}{-4}
\renewcommand{\theequation}{S\arabic{equation}}
\renewcommand{\thetable}{S\arabic{table}}

% \documentclass[aps,prb,reprint,onecolumn,natbib,showpacs,floatfix,superscriptaddress, longbibliography]{revtex4-2}

% \usepackage{graphicx}
% \usepackage{amsmath,amsthm,amssymb,amsfonts,bbold,bm,color,mathtools}
% \usepackage{float}
% \usepackage{epstopdf}
% \usepackage{hyperref}
% %\usepackage[monochrome]{color}
% \usepackage{enumerate}
% \usepackage{wasysym}
% \usepackage{url}
% \usepackage{dsfont}
% \usepackage{braket}
% \usepackage{comment} 
% \usepackage[capitalize]{cleveref}
% \usepackage[normalem]{ulem}
% \usepackage{xcolor}
% \usepackage[caption=false]{subfig}
% \graphicspath{{figures/}}
% \hypersetup{colorlinks=true,linkcolor=blue,citecolor=blue,urlcolor=blue}

% \begin{document}
% \newcommand{\pv}[1]{\textcolor{red}{#1}}
% \newcommand{\jt}[1]{\textcolor{orange}{#1}}
% \newcommand{\brac}[1]{\langle #1 \rangle}
% \newcommand{\para}[1]{\left( #1 \right)}

% \title{Supplementary Material}

% \author{Jefferson Tang}
% \affiliation{Department of Physics, University of Connecticut, Storrs, Connecticut 06269, USA}
% \author{Pavel A. Volkov}
% \affiliation{Department of Physics, University of Connecticut, Storrs, Connecticut 06269, USA}
% \begin{abstract}

% \end{abstract}
% \maketitle

\section{Current fluctuation in the presence of disorder}
Here we present the details of the calculation leading to Eq. \eqref{eq:disord}.
\begin{equation}
\begin{gathered}
        \langle \left[I^{(1)}\right]^2 \rangle_{\eta(x)}
        = 
    \left\langle
        \left(
        \int d\vec{r} 
    J_{c1}
    \cos(q_-x+\eta_{12}(\vec{r}))\sin\Big(\phi_0 + q_H x\Big)
    \right) \left(
        \int d\vec{r}' 
    J_{c1}
    \cos(q_-x'+\eta_{12}(\vec{r}'))
    \sin\Big(\phi_0 + q_H x'\Big)
    \right)
    \right
    \rangle_{\eta(x)} =
    \\
    \frac{J_{c1}^2}{4} \int 
    d \vec{r} 
    d \vec {r}'
    e^{-\frac{|\vec{r}-\vec{r}'|^2}{\xi^2}}
    \cos(q_- (x-x'))
    \left[
    \cos\Big(q_H (x-x')\Big)
    -
    \cos\Big(2\phi_0 + q_H (x'+x)\Big)
    \right]\approx
    \\
    \left. \approx \right|_{q_H \approx \pm q}
     \pi \xi^2 L W
     \frac{J_{c1}^2}{4}
     e^{-\frac{\xi^2 (|q_H|-q)^2}{4}}
\end{gathered}
\end{equation}

\section{Circular geometry}
\begin{figure}[h]
    \centering
    \includegraphics[width=0.5\linewidth]{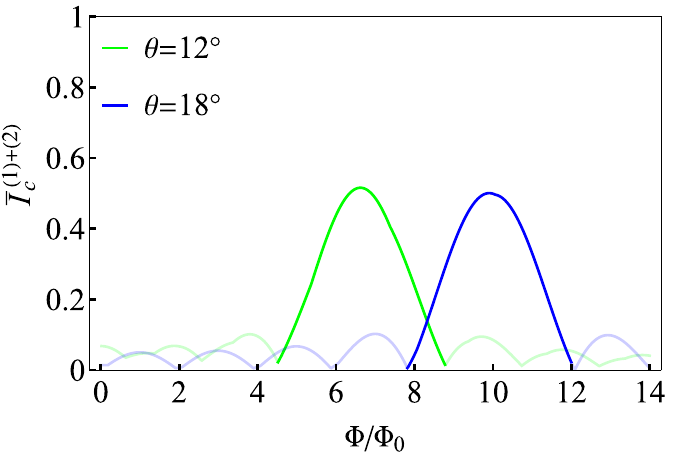}
    \caption{Critical Josephson current for a circular junction. We take $L/\lambda_{PDW}\approx 32$. }
    \label{fig:circular_junction}
\end{figure}

To obtain the Josephson current for a circular junction, we integrate Eq. \eqref{first_harmonic} , ignoring the $q_+$ term for reasons discussed in the main text, over a disk of radius $R$
\begin{equation}
     I^{(1)} = J_{c1}\int_{-R/2}^{R/2} dx\ \int_{-\sqrt{R/2-x}}^{\sqrt{R/2-x}}
    \cos(q_-x+\eta_{12})
    \sin\Big(q_Hx+\phi_0\Big)
\end{equation}
The results are shown in Fig. \ref{fig:circular_junction}.

\section{Josephson current from the coexistence of CDW and uniform superconductivity}

Inserting Eq. \eqref{phase_dif_mag} and Eq. \eqref{cdw_op} into Eq. \eqref{eq:I1} we integrate over a junction of length $L$ and width $W$ and get 
\begin{align}
    I^{(1)}_{\text{CDW+Uniform}} &= \int d\vec{r} \ \sin(q_Hx+\phi_0)\left[J_{c1}+J_{c2}\cos\left(k_-x+\varphi_{12}\right)\right] \nonumber \\
    &= J_{c1}\sin(\phi_0)\text{sinc}\left(\frac{\Phi}{2\Phi_0}\right)+J_{c2}\bar{I}^{(1)}\left(\frac{\Phi}{\Phi_0},\eta_{12},\phi_{12}\right)
\end{align}
where $\bar{I}^{(1)}$ is defined by Eq. \eqref{rect_current} and Eq. \eqref{eq:phimax}. \pv{For the PDW induced by coexistence, $J_{c2}/J_{c1} = \beta^2 |\rho_{k}|^2$.}

% \bibliography{ref}

% \end{document}

\end{document}